\documentclass[aps,prd,reprint,groupedaddress,showpacs,floatfix]{revtex4-1}

\usepackage{graphicx}
\usepackage{url}
\usepackage[letterpaper]{geometry}

\usepackage[utf8]{inputenc}

\newcommand{\tm}{\ensuremath{\tau^-}}
\newcommand{\tp}{\ensuremath{\tau^+}}
\newcommand{\tpm}{\ensuremath{\tp \tm}}
\newcommand{\Htpm}{\ensuremath{\mathrm{H \rightarrow \tpm}}}

\newcommand{\sth}{\ensuremath{\sin \theta^+ \sin \theta^-}}
\newcommand{\cthth}{\ensuremath{ c ( \theta^+ , \theta^- )}}

\newcommand{\zh}{\ensuremath{\mathrm{e^+ e^- \rightarrow H Z}}}
\newcommand{\taupi}{\ensuremath{\tau^\pm \rightarrow \pi^\pm \nu}}
\newcommand{\taurho}{\ensuremath{\tau^\pm \rightarrow \pi^\pm \pi^0 \nu}}
\newcommand{\psicp}{\ensuremath{\psi_\mathrm{CP}}}
\newcommand{\invab}{\ensuremath{\mathrm{ab^{-1}}}}

\newcommand{\Dphi}{\ensuremath{\Delta \phi}}
\newcommand{\dsm}{\ensuremath{d_\mathrm{sig}^\mathrm{min}}}
\newcommand{\elpr}{\ensuremath{\mathrm{e^-_L e^+_R}}}
\newcommand{\erpl}{\ensuremath{\mathrm{e^-_R e^+_L}}}
\newcommand{\erpr}{\ensuremath{\mathrm{e^-_R e^+_R}}}
\newcommand{\elpl}{\ensuremath{\mathrm{e^-_L e^+_L}}}

\begin{document}

\title{Measuring the CP state of tau lepton pairs from Higgs decay at the ILC}


\author{D. Jeans}
\email[]{daniel.jeans@kek.jp}
%\thanks{This study was performed in the framework of the ILD concept.}
\affiliation{Institute of Particle and Nuclear Studies, \\
High Energy Accelerator Research Organization (KEK), \\
Tsukuba, Japan.}

\author{G. W. Wilson}
\affiliation{University of Kansas, Department of Physics and Astronomy,
Malott Hall, 1251 Wescoe Hall Drive, Lawrence, KS 66045-7582, USA.
}

\collaboration{Study performed in the framework of the International Large Detector concept}

\date{\today}

\begin{abstract}%
In the Standard Model, the Higgs boson is a CP even state with CP conserving couplings; any deviations from 
this would be a sign of new physics.
These CP properties can be probed by measuring Higgs decays to $\tau$ lepton pairs:
the transverse correlation between the $\tau$ spins depends on CP.
This paper develops such an analysis, using full simulation of signal and background events in 
the International Large Detector concept for the International Linear Collider.
We consider Higgs-strahlung events (\zh) in which the Z boson decays to electrons, muons, or hadrons,
and the Higgs boson decays to $\tau$ leptons, which then decay either to $\taupi$ or $\taurho$. 
Assuming 2~\invab\ of integrated luminosity at a center-of-mass energy of 250~GeV, the mixing angle %$\psicp$ 
between even and odd CP components of the $\tau$ pair from Higgs boson decays can be measured to a precision
of 75~mrad ($4.3^\circ$).
\end{abstract}


\maketitle

\section{Introduction}

Several electron-positron collider designs are now being studied, whose major aim is to measure the Higgs sector with high precision,
thereby searching for effects of new physics beyond the Standard Model (SM)~\cite{ilctdr, CLIC_CDR, FCCee, CEPC}.
The inherently clean environment of lepton collisions and the high precision detectors possible at such
colliders will enable such precision measurements.
The most mature of these designs, 
the International Linear Collider~\cite{ilctdr, ILC250accel} (ILC), is a linear electron-positron collider with polarized beams,
which will initially operate at a center-of-mass energy of 250~GeV, with possible later energy upgrades to 500~GeV and potentially 1~TeV.
The wide range of precision Higgs sector measurements which can be performed at the ILC is, for example, summarized in~\cite{higgs_white_paper, ILC250phys}.
Key properties of the Higgs sector to be probed at such colliders are the Higgs boson mass, the strength of its coupling to other particles,
and its CP nature.

In the SM, the Higgs boson is a CP even scalar, while many extensions of the SM introduce additional Higgs bosons, often 
including a CP odd pseudoscalar state; Higgs boson mass eigenstates could be mixtures of such even and odd CP states.
In the SM, couplings of the Higgs boson to bosons and fermions are CP conserving, however additional terms inducing CP violation can be added to the 
Lagrangian, providing an additional potential source of non-SM CP effects.
A non-SM CP nature of the Higgs sector would produce several effects measurable at the ILC~\cite{higgs_white_paper},
including changes in the evolution of the \zh\ cross-section near threshold, and of the spin correlations between Higgs boson decay products
in boson or fermion decays. In many models of physics beyond the SM, CP-odd components of the Higgs boson do not couple directly to 
the W and Z bosons. The coupling to leptons is typically not suppressed, and therefore provides a more model independent approach
to probe the Higgs sector's CP properties.
In this paper we use the correlation between the spins of $\tau$ leptons produced in Higgs boson decay to study CP properties of the Higgs sector.

The $\tau$ lepton provides a powerful tool with which to probe the CP properties of the Higgs boson.
The branching ratio of the Higgs boson to \tpm\ is relatively large ($\sim 6.3\%$ in the SM), 
and the spin correlations of the two $\tau$ leptons, on which the measurement relies, 
are not affected by strong interactions in the final state.
The mean lifetime of the $\tau$ lepton, $\mathrm{87~\mu m/c}$, is short enough to allow its decay products to be measured in the detector, 
providing access to the the spin of the $\tau$ lepton, yet long enough 
to allow impact parameters measured by a vertex detector to be used in its reconstruction.

This paper demonstrates how the CP properties of $\tau$ lepton pairs produced in Higgs boson decays can be measured, 
and estimates the precision that can be achieved by the ILC operating at 250~GeV.
Signal and background processes were fully simulated and reconstructed in the International Large Detector (ILD) concept~\cite{detectors_dbd}.
The signal was considered to be the Higgs-strahlung process (\zh), with Z decays into electrons, muons, and hadrons.
Other Z decays (into $\tau$s or neutrinos) are less useful due to incomplete detection of the Z decay products.
Events in which the Higgs boson decays to a pair of $\tau$ leptons were analysed. The reconstruction method developed in \cite{jeans_taus}
allows the momenta of hadronically decaying $\tau$ leptons to be fully reconstructed by making use of the interaction point position, the impact parameters of the
$\tau$ lepton decay products, and the transverse momentum of the system recoiling against the \tpm\ system.
The $\taupi$ and $\taurho$ decay channels were considered, which respectively account for 11\% and 26\% of $\tau$ lepton decays.
These $\tau$ lepton decays allow full reconstruction of the $\tau$ lepton momenta and provide optimal sensitivity to the direction of the $\tau$ spin.
The optimal estimator of the $\tau$ lepton spin direction, or polarimeter, is extracted from the $\tau$ lepton decay products' momenta,
and the CP state of the $\tau$ pair is extracted by considering the correlation between components of the 
two polarimeters transverse to the $\tau$ lepton momenta.

The use of $\tau$ lepton spin correlations to probe the CP nature of the Higgs boson has been investigated in several signal-only, 
generator or fast simulation level studies, see for example~\cite{Nelson,BargerZerwas,gunion,Was_transverse,desch,rouge,BergeILC,Harnik,Chen}.
A full simulation study at the ILC in one particular final state ($\zh, \mathrm{Z \rightarrow \mu^+ \mu^- }$)
has been presented in~\cite{marcel}.

The sensitivity of the LHC experiments to measure CP effects in \Htpm\ has been investigated in 
a number of phenomenological studies.
In \cite{BergeLHC2015} a precision on the CP mixing of $4^\circ$ is predicted with an integrated luminosity of $3~\mathrm{ab}^{-1}$,
while \cite{Harnik} and \cite{Hagiwara} both suggest a precision of $\sim11^\circ$ using the same integrated luminosity. 
In \cite{Askew} it is argued that the analysis of \cite{Harnik} is significantly affected by experimental effects,
and that even and odd CP hypotheses can be distinguished at no better than 95\% confidence level with an integrated
luminosity of $1~\mathrm{ab}^{-1}$.

The method used to extract CP-sensitive observables is described in Section~\ref{sec:cpang}.
The procedures used to generate, simulate, and reconstruct events are outlined in Section~\ref{sec:gensim}, 
and the event selection is described in Section~\ref{sec:selection}. 
Estimates of the measurement sensitivity at the ILC are presented in Section~\ref{sec:results}, 
and conclusions drawn in Section~\ref{sec:conclusion}.

\section{CP in \Htpm}
\label{sec:cpang}

The CP state of the $\tau$ lepton pair produced in Higgs boson decay is determined by the CP properties of both the Higgs mass eigenstate
and the $H\tpm$ vertex. In the SM, the Higgs boson is a CP even state, and its couplings conserve CP.
In more general models, a Higgs mass eigenstate $\mathrm{h_m}$ can be written in terms of CP even ($\mathrm{h}$) and odd ($\mathrm{A}$) components as 
\begin{equation}
\mathrm{h_m = h \cos \psicp + A \sin \psicp ,}
\label{eqn:masscpv}
\end{equation}
where $\mathrm{h_m}$ is purely CP even (odd) when $\psicp = 0 \ (\pi/2)$. 
The full range of \psicp\ is $0 \to 2 \pi$, however differences between \psicp\ and $\psicp+\pi$ will appear only in interference terms,
and will be very challenging to measure.
CP violation in the coupling can explicitly be induced by a Lagrangian term such as 
\begin{equation}
\mathcal{L}_\mathrm{H\tau\tau} = g \overline{\tau} ( \cos \psicp + i \gamma_5 \sin \psicp ) \tau \mathrm{H}
\label{eqn:cpv}
\end{equation}
which is CP-conserving %(and SM--like) 
for $\psicp = 0$ and maximally violates CP for $\psicp = \pi/2$. 
The goal of the present analysis is to determine the CP state of the $\tau$-pair,
parameterized in terms of %the angle 
$\psicp$. 
A non-zero value of \psicp\ in $\tau$-pairs from Higgs boson decay could arise from either or both of the above mechanisms.
These scenarios can be distinguished by considering other observables; for example a mixed CP Higgs mass eigenstate 
would affect the total Higgs-strahlung cross-section %(which relies on the $HZZ$ coupling), 
which would not be the case if non-SM CP effects occur only in the coupling to fermions.

\begin{figure}
\centering
\includegraphics[width=0.49\textwidth]{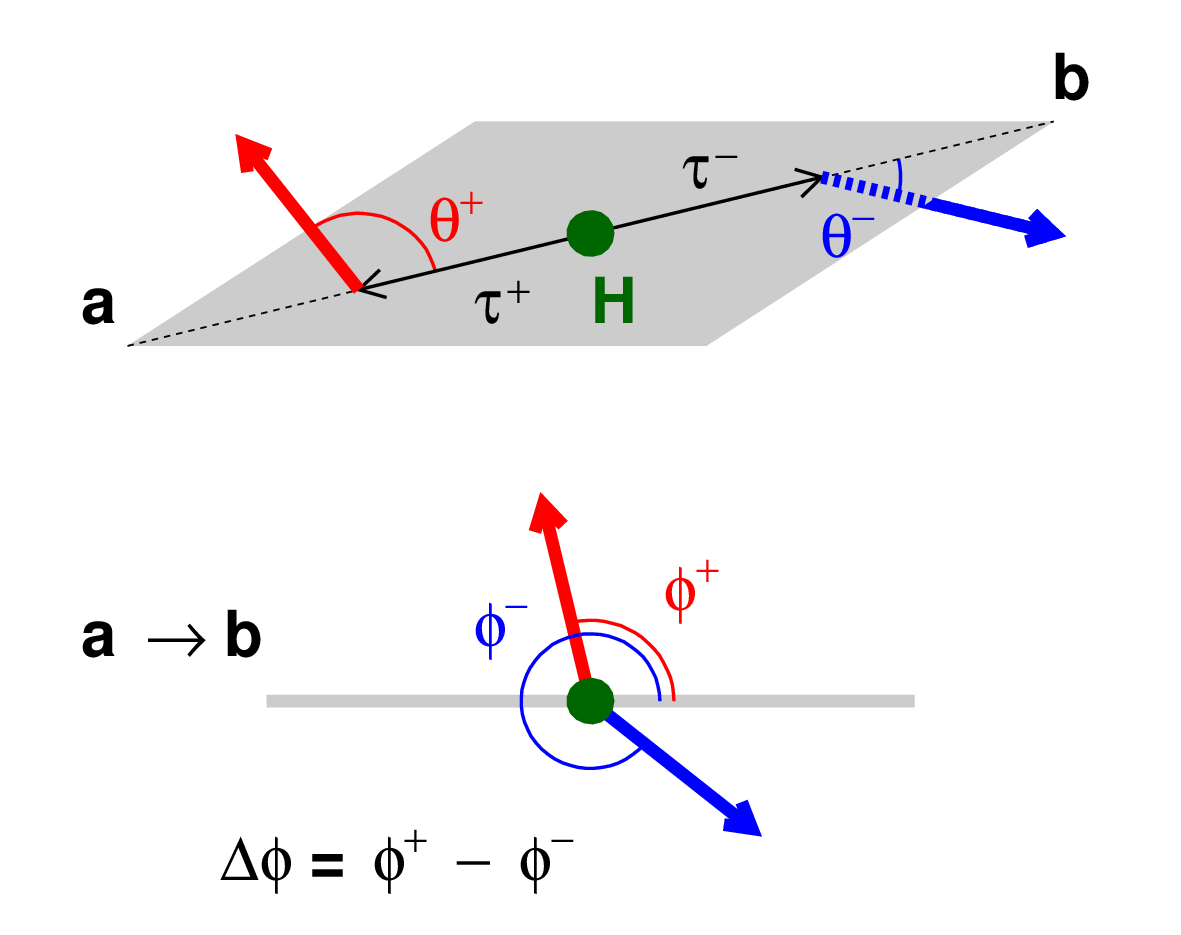}
\caption{Definition of the angles $\theta^\pm, \phi^\pm$. The $\tau^\pm$ momentum (polarimeter) vectors are shown as thin (thick) arrows.
The second picture shows the view looking along the line from a to b (i.e. along the direction of the \tm\ momentum). Angles are defined in the respective $\tau$ rest frames.
}
\label{fig:angles}
\end{figure}

The CP state of a \tpm\ pair produced in the decay of a scalar state $h$
affects the correlation between the components of the tau polarization
perpendicular to the $\tau$ momentum direction~\cite{BargerZerwas}.
The optimal estimator of the $\tau$ polarization is the so-called {\em polarimeter} (or {\em effective spin}) vector, which can be
reconstructed from the momenta of the $\tau$ lepton's decay products.
It is shown in~\cite{gunion} that the distribution of the $\tau$ polarimeter vectors in 
the decay of a scalar boson
to \tpm\ can be written as
\begin{eqnarray}
dN / ( d &&\cos \theta^+ d  \cos \theta^- d \phi^+ d \phi^- )  \nonumber\\
\propto \ && (b^2 + a^2 \beta_\tau^2)(1 + \cos \theta^+ \cos \theta^- ) \nonumber\\
 + && (b^2 - a^2 \beta_\tau^2)  \sth  \cos (  \phi^+ - \phi^- ) \nonumber \\
 - && 2 a b \beta_\tau \sth  \sin (  \phi^+ - \phi^- ) 
\label{eqn:orig}
\end{eqnarray}
where $\theta^\pm, \phi^\pm$ are respectively the polar and azimuthal angles of the polarimeter vector with respect 
to ($\pm$) the Higgs boson momentum direction as evaluated in the respective $\tau^\pm$ rest frames, as
illustrated in Fig.~\ref{fig:angles}, and
$\beta_\tau$ is the $\tau^\pm$ velocity in the $\tpm$ rest frame.
The factors $(a, b)$ of Eqn.~\ref{eqn:orig} define \psicp: $\tan(\psicp) \equiv b/a$. Defining $\Delta \phi \equiv  \phi^+ - \phi^-$,
we rewrite Eqn.~\ref{eqn:orig} as
\begin{eqnarray}
 dN / ( d \cos &&\theta^+ d  \cos \theta^- d \phi^+ d \phi^- )  \nonumber\\
\propto \ && ( \sin^2 \psicp +  \beta_\tau^2 \cos^2 \psicp )  \nonumber\\
          && \ \times ( 1 + \cos \theta^+ \cos \theta^- ) \nonumber\\
        + && ( \sin^2 \psicp - \beta_\tau^2 \cos^2 \psicp )  \nonumber\\
          && \ \times \sth \cos ( \Dphi ) \nonumber\\
        - && 2 \beta_\tau \cos \psicp  \sin \psicp \nonumber\\
          && \ \times \sth \sin ( \Dphi ).
\end{eqnarray}

Since the $\tau$ leptons produced in the decay of a 
$125~\mathrm{GeV/c^2}$ Higgs boson are highly relativistic, we take the limit $\beta_\tau \to 1$: 
\begin{eqnarray}
dN / ( d &&\cos \theta^+ d  \cos \theta^- d \phi^+ d \phi^- )  \nonumber\\
 \propto && \ 1 + \cos \theta^+ \cos \theta^- \nonumber\\
         && - \sth \cos ( \Dphi - 2 \psicp ) \nonumber\\
 \propto && \ ( 1 + \cos \theta^+ \cos \theta^- ) \nonumber\\
         && \times \big( 1 - \cthth \cos ( \Dphi - 2 \psicp ) \big),
\label{eqn:master}
\end{eqnarray}
where we define the contrast function $\cthth \equiv \sth / ( 1 + \cos \theta^+ \cos \theta^- )$.

\begin{figure}
\centering
\includegraphics[width=0.49\textwidth]{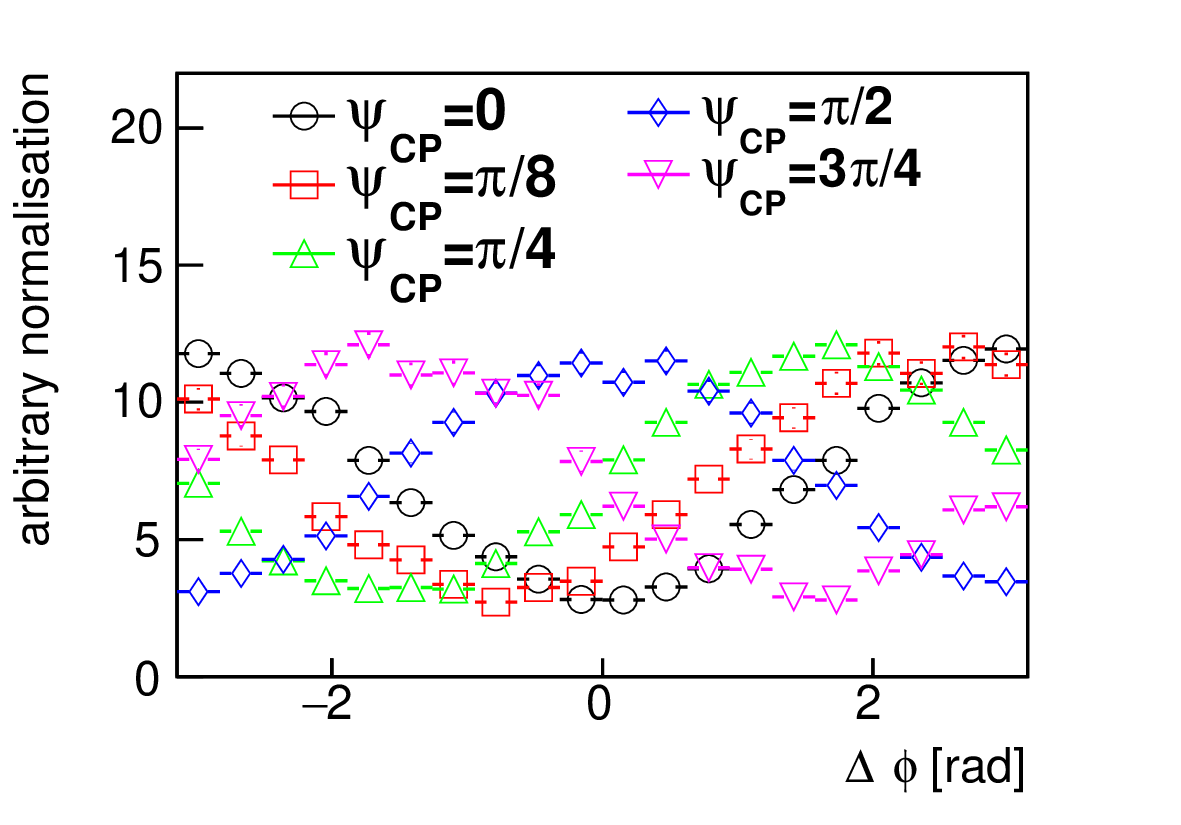} 
\caption{
Distributions of events in \Dphi, integrated over $\theta^\pm$, at MC truth level, for different values of \psicp.
}
\label{fig:mc_dphi}
\end{figure}

\begin{figure}
\centering
\includegraphics[width=0.49\textwidth]{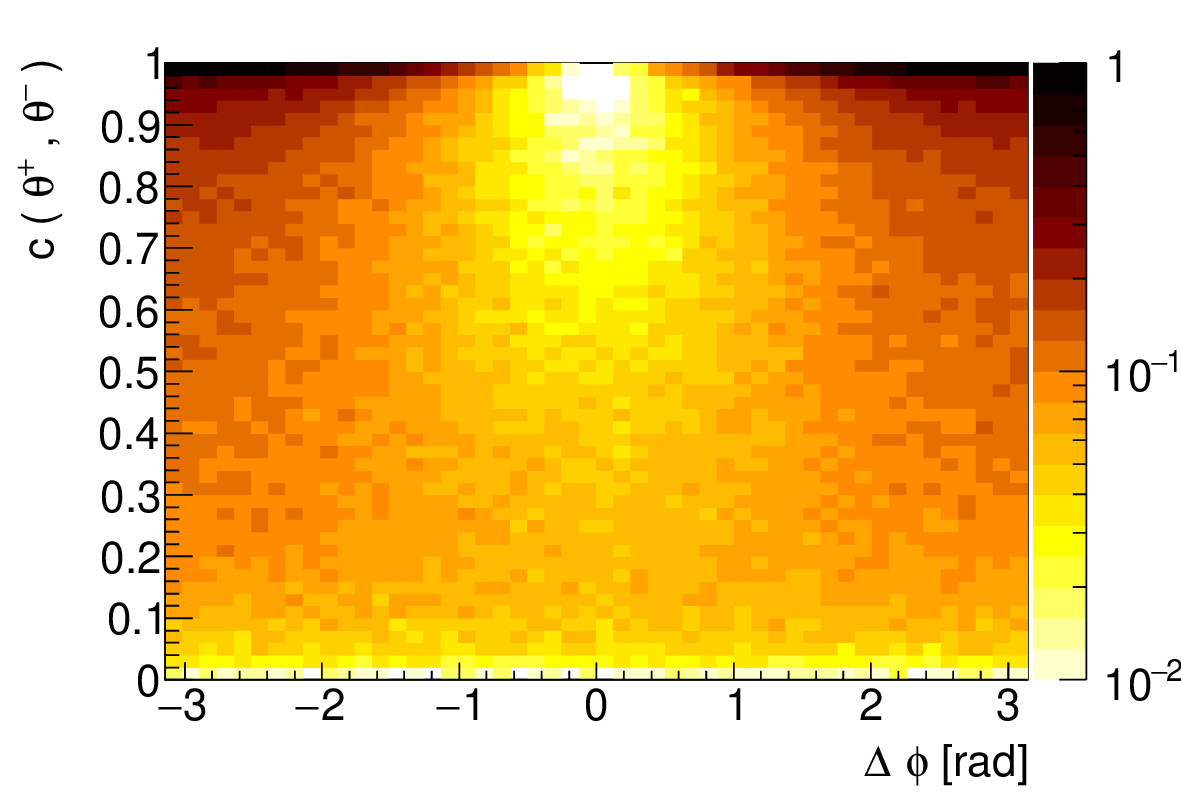}
\caption{
Two-dimensional distribution of events in \Dphi\ and \cthth\ at MC truth level, for the case $\psicp=0$.
}
\label{fig:mc_dphi_cont}
\end{figure}

It can be seen from Eqn.~\ref{eqn:master} that \psicp\ affects the distribution of events in \Dphi, and
that the strength of this effect in a particular event depends on $\theta^\pm$ via the contrast function $\cthth$.
Figure~\ref{fig:mc_dphi} shows, at Monte Carlo (MC) truth level, the \Dphi\ distribution for different values of \psicp,
while the dependence of the $\Delta \phi$ distribution on $\cthth$ is shown in Fig.~\ref{fig:mc_dphi_cont}.

Polarimeter vectors $\mathbf{h}$ are straightforward to calculate in the $\taupi$ and $\taurho$ decay modes. 
Using the conventions in~\cite{gunion}, they can be written in the $\tau$ rest frames as
\begin{eqnarray}
\mathbf{h} ( \taupi )  \propto \ && \mathbf{p}_{\pi^\pm} \\
\mathbf{h} ( \taurho ) \propto \ && m_\tau ( E_{\pi^\pm} - E_{\pi^0} ) ( \mathbf{p}_{\pi^\pm} - \mathbf{p}_{\pi^0}) \nonumber\\
                       && + \frac{1}{2} ( p_{\pi^\pm} + p_{\pi^0} )^2 \mathbf{p}_\nu ,
\label{eqn:polar}
\end{eqnarray}
where $p_{\pi^\pm}$, $p_{\pi^0}$, $p_\nu$ are respectively the four-momenta of the charged and neutral pions, and of the neutrino.

The strategy of the present analysis is to reconstruct the \Dphi\ distribution in \zh, \Htpm\ events, and to use this 
distribution to extract the value of \psicp. 

\section{Event generation, simulation, and basic reconstruction}
\label{sec:gensim}

The {\sc whizard} event generator (version 2.2.8)~\cite{whizard} was used to generate
$\mathrm{e^+ e^- \rightarrow f \overline{f} \tpm}$ events (where $\mathrm{f = e, \mu, u, d, s, c, b}$).
{\sc circe2}~\cite{circe2} was used to model the beam spectrum of the ILC at a center-of-mass energy of 250 GeV,
and the effects of initial state radiation were simulated.
For each final state, one sample was produced imposing that the \tpm\ was produced 
in the decay of a Higgs boson ($\mathrm{m_H = 125~GeV/c^2}$), 
and a second sample without the Higgs contribution.
{\sc pythia} (version 8.212)~\cite{pythia6, pythia8} was used to model Final State Radiation (FSR), hadronize quarks, and decay $\tau$ leptons.
Longitudinal and transverse spin correlations between the two $\tau$ lepton decays were included~\cite{pythia_taus}. 
Different correlations were applied to $\tau$ lepton pairs from H decay and those from $Z/\gamma$, as appropriate. 
Two sets of samples were prepared, the first including only the $\taupi$ and $\taurho$ decay channels,
and the second with all $\tau$ decays. Samples with varying Higgs CP properties were simulated by 
changing the spin correlations applied in the decay of the $\tau$ pair by means of {\sc pythia}'s {\tt HiggsH1:phiParity} parameter
to describe $\psicp=0$ (i.e. the SM), $\pi/8, \pi/4, \pi/2$, and $3\pi/4$~rad.
The effective integrated luminosity of the simulated signal samples was significantly larger than the expected
ILC integrated luminosity.

A full set of 2- and 4-fermion backgrounds from other SM processes was considered. 
These samples were centrally generated by the ILC physics group using {\sc whizard} version 1.96.
The effective integrated luminosity generated for some of these processes, in particular those with high cross-section, 
was smaller than that expected at ILC.

Samples were produced with two pure initial polarization states, 
(left-handed $\mathrm{e^-}$, right-handed $\mathrm{e^+}$) and 
(right-handed $\mathrm{e^-}$, left-handed $\mathrm{e^+}$), which
were mixed with appropriate weights to simulate the proposed mix of polarization states at ILC.
The electron (positron) beam will have 80\% (30\%) polarization.
In this analysis we assume a total integrated luminosity of 2~\invab\ at a center-of-mass energy of 250~GeV, 
distributed among the different polarization sign combinations
[\elpr, \erpl, \elpl, \erpr] 
as [45, 45, 5, 5]\%, 
corresponding to the 250~GeV portion of the ``H20-staged'' ILC running scenario proposed in~\cite{ILC250phys}.
The \elpr\ portion, for example, has a dominantly left-handed electron beam 
and dominantly right-handed positron beam.

Events were simulated in the International Large Detector (ILD)~\cite{detectors_dbd}, a detector concept for the ILC.
It consists of a high precision silicon vertex detector, a large time projection chamber, 
additional silicon strip tracking detectors, and
highly granular electromagnetic, hadronic, and forward calorimeters, 
all placed within the 3.5~T field of a solenoid whose iron flux return yoke is instrumented with muon detectors.
Simulation of the {\tt ILD\_o1\_v05} detector model~\cite{detectors_dbd} was performed using the {\sc geant4}-based {\sc mokka} toolkit~\cite{mokka}.
Background due to the interaction of beam remnants was superimposed on the simulated events.
The simulated energy deposits were digitized and reconstructed using {\sc marlinreco} and other packages of {\sc ilcsoft}~\cite{ilcsoft} 
(version v01-16-02).
The output of event reconstruction is a collection of Particle Flow Objects (PFOs, 
each containing zero, one or possibly more reconstructed tracks and calorimeter clusters), 
corresponding to individual final state particles.

ILD's charged track impact parameter resolution in the $x-y$ plane is at least as good as the stated goal of
$\sigma_{d_0} \sim 5~\mu m \oplus 10~\mu m / ( p [\mathrm{GeV/c}] \sin^{3/2} \theta)$,
while in the $r-z$ plane the impact resolution is better than $\sim 10~\mu m$ for track momenta above 3~GeV/c~\cite{detectors_dbd}.
The highly granular readout of the electromagnetic calorimeter allows nearby photons to be efficiently resolved, 
leading to excellent identification of $\tau$ lepton decay modes~\cite{Tran:taus}.

Simple particle identification was applied to PFOs, based on the amount and distribution of energy deposits in the calorimeters.
Charged PFOs were classified as either electrons, muons, or hadrons, while neutral PFOs were classified as either photons or neutral hadrons.
Photon PFOs close to charged particles were considered as being induced by final state radiation (FSR) or bremsstrahlung.

\section{Event selection and reconstruction}
\label{sec:selection}

This section describes the methods used to reconstruct and select signal events, 
while rejecting background events produced by other processes.
Signal events are defined as Higgs-strahlung events in which: 
the Z decays to either electrons, muons, or quarks; 
the Higgs boson decays to a $\tau$ lepton pair; and 
both $\tau$s decay to either $\taupi$  or  $\taurho$. 
The method used to fully reconstruct the $\tau$ lepton momenta is not applicable to Z decays to neutrinos or $\tau$ leptons, since
neither the production vertex nor the recoiling momentum can be precisely determined. The extraction of the $\tau$ lepton
spin information is more complex in the other $\tau$ lepton decay modes.

\subsection{Preselection for Z decays to electrons or muons}

Events with less than eight charged PFOs were considered in the leptonic (electron or muon) selection channel.
The Z decay into electrons or muons was first identified, and $\tau$s were searched for in the remainder of the event.
Charged PFOs with a reconstructed energy of at least 12~GeV (when combined with any identified bremsstrahlung and FSR photon PFOs) 
were used to search for Z candidates.
A pair of oppositely charged PFOs with an invariant mass within $\mathrm{20~GeV/c^2}$ of the Z boson mass was 
considered a Z decay candidate; at least one of the pair was required to be identified as an electron or muon, 
and the two could not be identified as differently flavored leptons.
Identified electrons were required to have 
$|\cos ( \theta)| < 0.95$ 
to reduce backgrounds due to e.g. $\mathrm{We\nu}$ and $\mathrm{Zee}$ final states
in which the final state $\mathrm{e^\pm}$ is often in the very forward region.
If no Z candidate was identified, the event was rejected, while if more than one was found, 
the one with invariant mass closest to the Z mass was retained.

A search was made for a pair of single-prong hadronic $\tau$ decays among the PFOs not assigned to the Z.
The two most energetic charged PFOs were considered $\tau$ decay prongs. 
These two PFOs were required to have opposite charge, and an estimated uncertainty on their impact parameter measurement no larger than $25~\mu m$.
For each prong, additional charged PFOs within $10^\circ$ were considered: if there was more than one such PFO, or if the sum of their energies was
greater than 3~GeV, the event was rejected.

Photon PFOs not associated with the Z decay were then considered for inclusion in the $\tau$ jets.
Starting from the most energetic photon PFO, photon pairs consistent with the $\pi^0$ mass 
(based on the result of a constrained kinematic fit in which photon energies were varied within their expected measurement uncertainties 
while imposing the $\pi^0$ mass), 
and which when combined with an existing $\tau$ jet did not increase its mass over $m_\tau$, were assigned to the $\tau$ jets.
In a second step, unassigned photon PFOs were added to the closest $\tau$ jet if the resulting system's invariant mass did not exceed $m_\tau$,
starting with the highest energy photon candidates.

If the visible invariant mass of a $\tau$ jet was less than $\mathrm{0.2~GeV/c^2}$, and the prong was not identified as an electron or muon, 
it was considered a $\taupi$ decay. If the invariant mass was between 0.2 and 1.2~$\mathrm{GeV/c^2}$, it was considered a $\taurho$ decay.
If either $\tau$ jet passed neither of these criteria, the event was rejected.

\subsection{Preselection for hadronic Z decays}

Events with at least eight charged PFOs were considered in the the hadronic channel. 
A search was made for isolated single-prong jets, excluding PFOs in the very forward region ($|\cos \theta|>0.95$) to reduce
contamination from beam-related backgrounds.
Charged PFOs were considered isolated if the isolation angle, defined as the angle to the nearest charged PFO, was at least $6^\circ$.
Isolated charged PFOs were ranked according to the product of their momentum and isolation angle; those
with a score below 0.8 (GeV/c deg.) were rejected.
If less than two isolated prongs survived, the event was rejected. 
The highest-ranked pair of oppositely-charged isolated PFOs was used to form two tau jet seeds.
If either of these PFOs had an estimated impact parameter uncertainty larger than $25~\mu m$, the event was rejected.

Photons within $10^\circ$ of a tau jet seed were considered for inclusion into its jet.
A similar method was used as in the leptonic channel, based on matching photons into $\pi^0$s using a kinematic fit, 
and on the resulting jet's invariant mass.
The tau jets were required to be consistent with a $\taupi$ or $\taurho$ decay, using the same criteria as used in the leptonic channel.

PFOs associated with neither of the $\tau$ jets were assigned to the system recoiling against the pair of $\tau$ leptons.
The invariant mass of this recoiling system (corresponding to the Z decay products in the signal process) was required to be between 60 and 160 GeV, 
and the mass recoiling against it (corresponding to the Higgs boson mass) was required to be between 50 and 170 GeV.

\subsection{Event reconstruction}

Charged PFOs associated with the Z boson decay were used to reconstruct the primary vertex (PV).
In the case of more than two such PFOs, an iterative procedure was used to prune this vertex,
repeatedly removing the track which contributed most to the $\chi^2$ of the vertex, if this contribution exceeded 10. 
The position of this pruned vertex was considered the point of $\tau$ production.
The size of the ILC interaction region is expected to be approximately 
$\sigma_{x \times y \times z} \sim {\mathrm 1~\mu m \times 8~nm \times 300~\mu m}$.
The size in $z$ is significantly larger than typical PV position resolution, however the small size in $x-y$ could provide a 
useful additional constraint on the PV position. This additional information was not, however, used in this analysis.

\begin{figure*}
\centering
\includegraphics[width=0.49\textwidth]{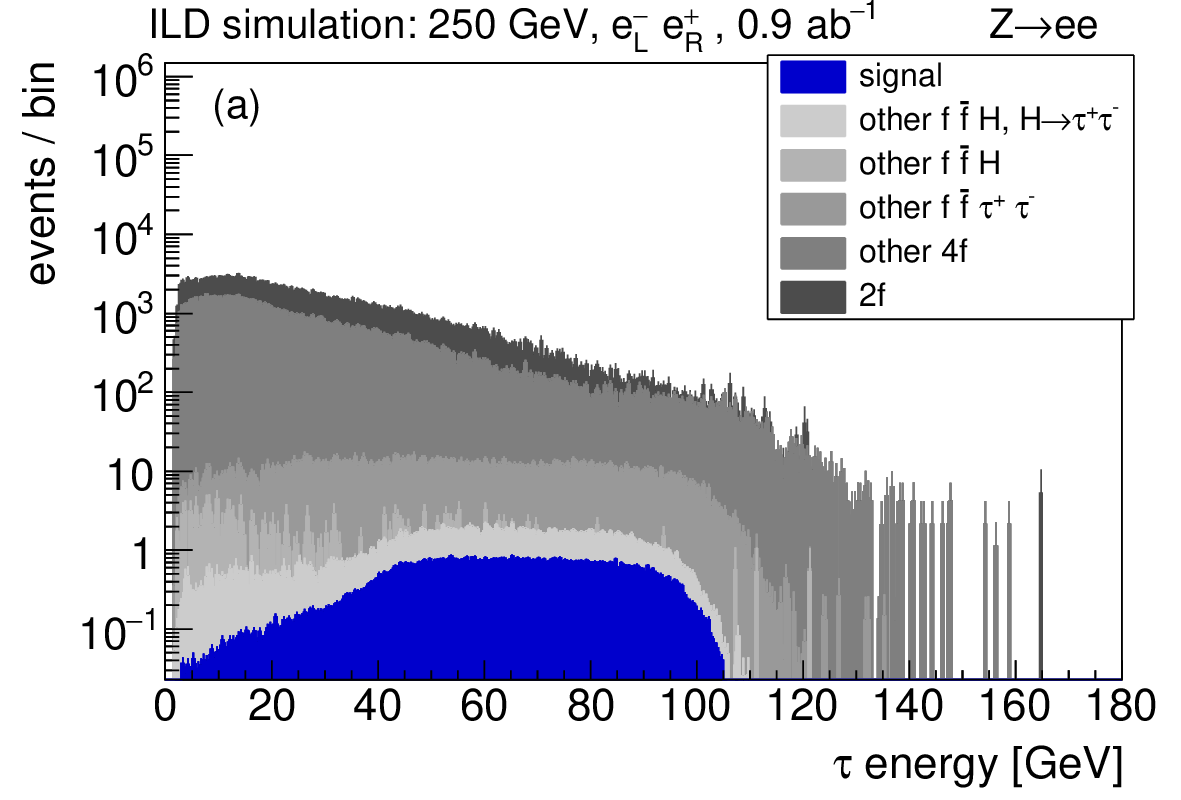}
\includegraphics[width=0.49\textwidth]{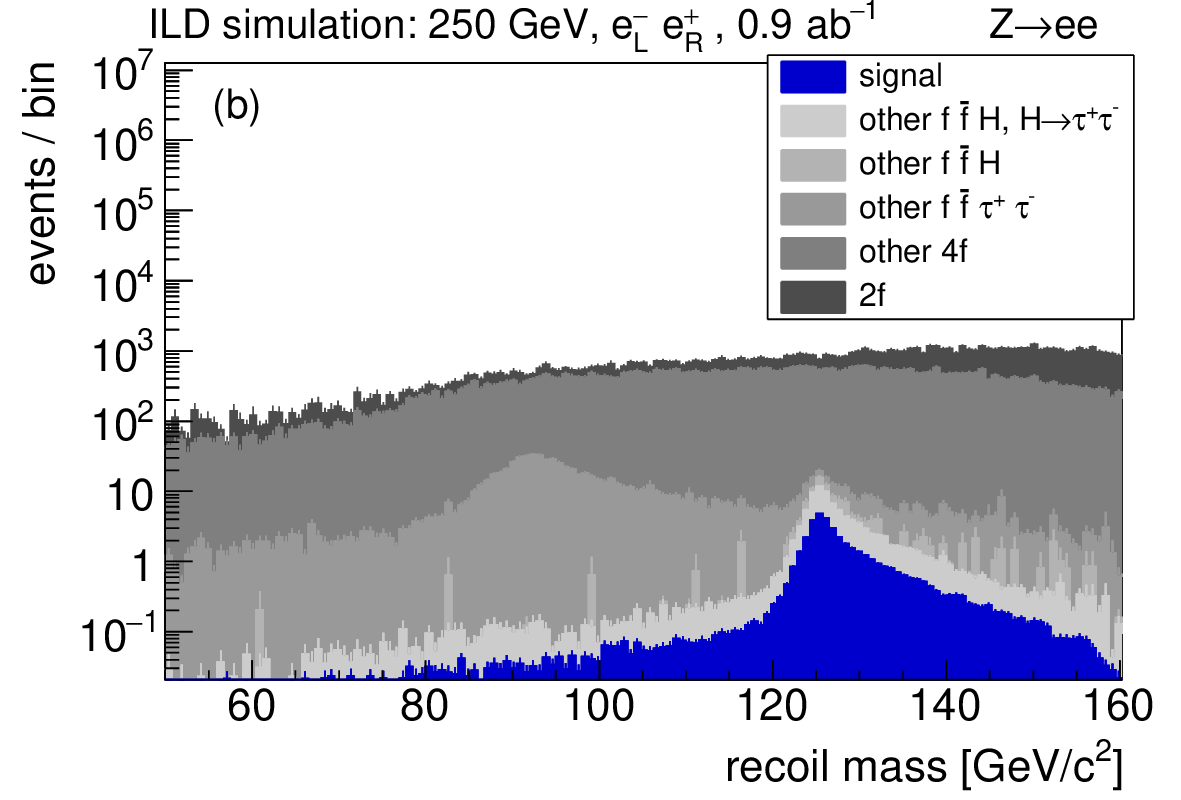}\\
\hspace{0.5cm}\\
\includegraphics[width=0.49\textwidth]{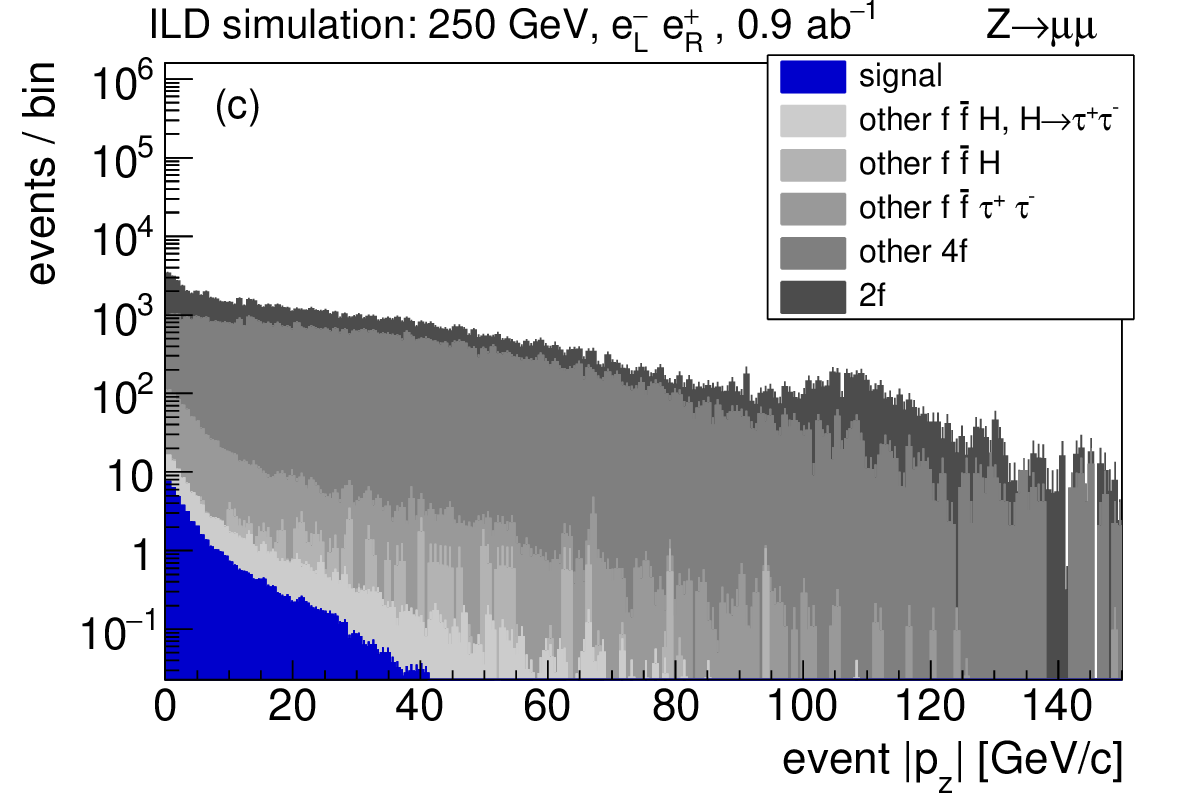}
\includegraphics[width=0.49\textwidth]{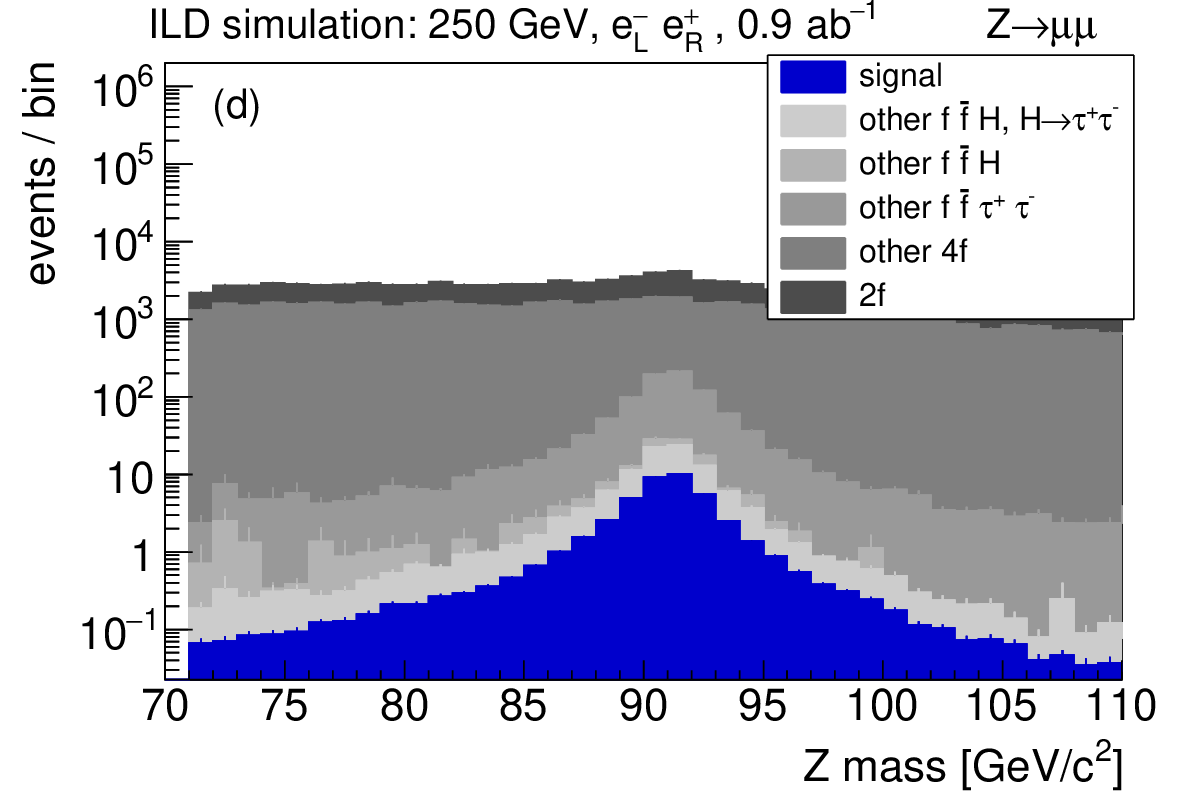}\\
\hspace{0.5cm}\\
\includegraphics[width=0.49\textwidth]{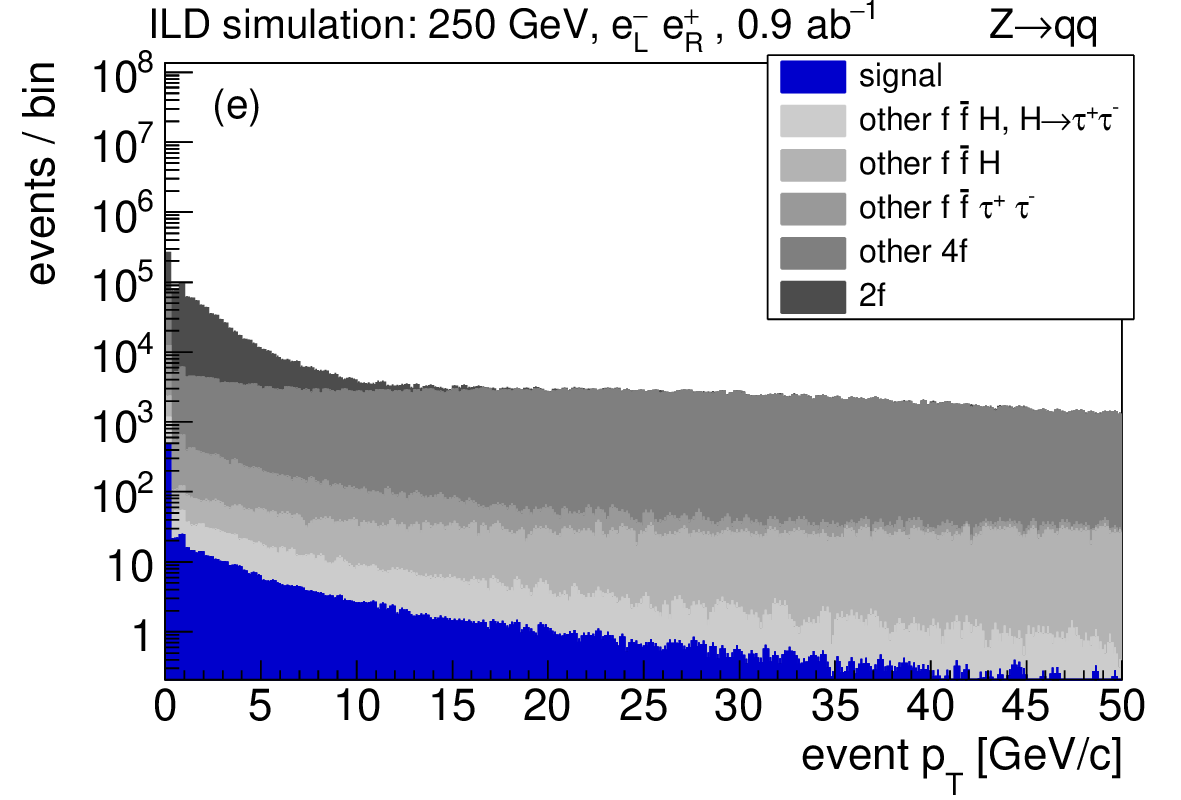}
\includegraphics[width=0.49\textwidth]{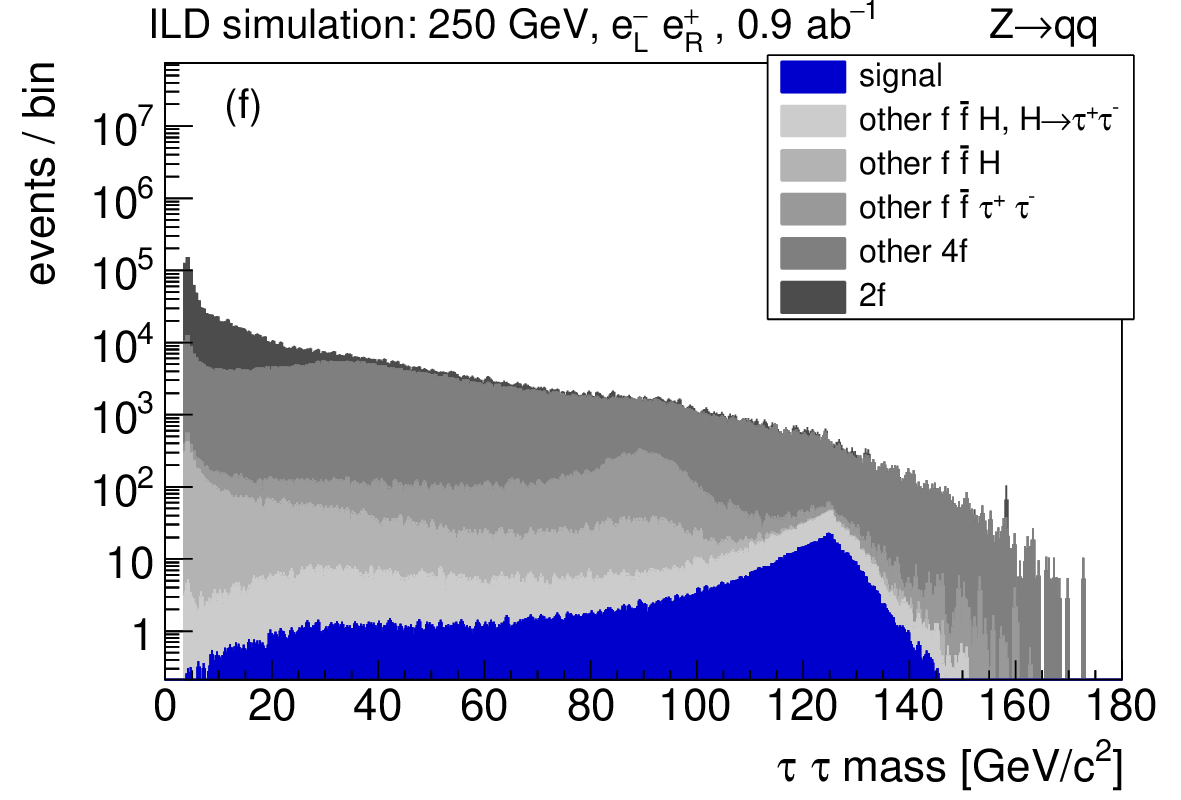}
\caption{Distributions of some reconstructed observables after preselection and reconstruction in different 
selection channels:
(a) the reconstructed $\tau$ lepton energy, 
(b) the mass recoiling against the system assigned to the Z boson, 
(c) the net momentum of the event in the beam direction, 
(d) the invariant mass of the system assigned to the Z boson,
(e) the ${\mathrm p_T}$ of the event, and 
(f) the invariant mass of the reconstructed $\tau$ lepton pair.
Distributions are normalized to 0.9~\invab\ of data in the \elpr\ beam polarization.
}
\label{fig:presel_vars}
\end{figure*}

The measured properties of events were used to fully reconstruct the $\tau$ momenta, using the method developed in~\cite{jeans_taus}.
The system of two hadronically-decaying $\tau$ leptons has six unmeasured parameters, corresponding to the three-momenta
of the two neutrinos produced in the $\tau$ decays.
These six parameters can be determined, up to two-fold ambiguities, by the use of six constraints which are applicable
to the present analysis.
Each $\tau$ lepton's momentum is constrained to lie in the plane which contains the PV and the tangent to the charged 
$\tau$ daughter's trajectory at it's closest approach to the PV.
In addition, the total invariant mass of each $\tau$'s decay products, including the neutrino, is constrained to be $\mathrm{1.777~GeV/c^2}$.
These four constraints per $\tau$ pair leave a single free parameter per $\tau$ lepton, which it is convenient to express as the angle between the components of the neutrino 
and hadronic momenta in the aforementioned $\tau$ momentum plane. 
These two angles, one per $\tau$, are then chosen to minimize the total transverse momentum of the event, 
including the two $\tau$s and the system recoiling against them.
Two-fold ambiguities, which arise from the quadratic invariant mass constraints, are resolved by rejecting solutions in which either $\tau$ lepton is reconstructed
with a negative decay length,
and, if ambiguities remain, by choosing the solution with invariant \tpm\ mass closest to the Higgs boson mass of $\mathrm{125~GeV/c^2}$.
Since this method relies on the balance of momentum transverse to the beamline, any boost along the beam direction (e.g. due to initial state radiation
or beamstrahlung) has no effect on the reconstruction of the $\tau$ lepton momenta.
Distributions of some event observables after preselection and reconstruction are shown in Fig.~\ref{fig:presel_vars}.

The efficiency to preselect and successfully reconstruct signal events (i.e. events in which the two $\tau$ leptons 
from a Higgs boson decay into one of the considered decay modes) is around 58\% for both leptonically and hadronically decaying Z bosons.
Table~\ref{tab:tautaudecmodes} demonstrates the performance of the $\tau$ decay mode identification in preselected and reconstructed events.
The efficiency to correctly assign the decay modes of both $\tau$s in an event is around 93\% and 88\% respectively in the 
leptonic and hadronic channels.

\begin{table}
 \caption{Migrations among $\tau$-pair decay modes, for preselected and reconstructed signal events in which the Z boson decays to either muons or light quarks. 
All numbers are given in \%.}
  \label{tab:tautaudecmodes}
  \centering
  \begin{ruledtabular}
  \begin{tabular}{lccc}
                &  \multicolumn{3}{c}{True decay} \\
Reco. decay                &   ($\pi\nu, \pi\nu $) & ($\pi\nu, \rho\nu$) & ($\rho\nu, \rho\nu$) \\
  \hline
  &  \multicolumn{3}{c}{$\mathrm{Z \to \mu^+ \mu^-}$}  \\
 ($\pi  \nu, \pi \nu$) & $93$    & $3 $  & $<1 $   \\
 ($\pi  \nu, \rho\nu$) & $7 $    & $93$  & $6  $   \\
 ($\rho \nu, \rho\nu$) & $<1$    & $4 $  & $94 $   \\
  \hline
  &  \multicolumn{3}{c}{$\mathrm{Z \to qq (uds)}$} \\
 ($\pi  \nu, \pi \nu$) & $89$  & $ 6  $  & $<1 $ \\
 ($\pi  \nu, \rho\nu$) & $11$  & $ 89 $  & $12 $ \\
 ($\rho \nu, \rho\nu$) & $<1 $  & $ 5 $  & $87 $ \\
 \end{tabular}
 \end{ruledtabular}
 \end{table}

\subsection{Event selection}

A rather loose selection was then applied to remove badly reconstructed signal events and the majority of background events.
Requirements were placed on 
the reconstructed $\tau^+ \tau^-$ invariant mass ($m_{\tau\tau}$), 
the polar angle of the least forward charged $\tau$ lepton decay prong $|\cos \theta_{P}|_\mathrm{min}$,
the invariant mass, polar angle, and mass recoiling against the system associated with the Z ($m_Z$, $|\cos \theta_Z|$, $m_\mathrm{recoil}$), 
and on the net event momentum in the transverse and $z$ directions ($p_T, p_z$), as shown in Table~\ref{tab:cuts_effs}.
The table also lists the selection efficiencies and remaining backgrounds at each step of the selection.

\begin{table*}
\caption{Selection cuts [see text for details; (energies, momenta, and masses) in $\mathrm{GeV/c^{(0,1,2)}}$], signal selection efficiencies $\epsilon$ (in \%), 
and number of expected background events ($\mathrm{BG}$)
at various stages of the selection in the three selection channels $e, \mu, q$. 
Event numbers are scaled to the 2~\invab\ of 250~GeV data of the ``H20-staged'' running scenario.
}
\label{tab:cuts_effs}
\centering
\begin{ruledtabular}
\begin{tabular}{llrrrrlrr}
                           & \multicolumn{5}{c}{leptonic preselection} & \multicolumn{3}{c}{hadronic preselection} \\ 
\cline{2-6} \cline{7-9} 
event property             & requirement &  $\epsilon_e$   &  $\epsilon_\mu$    & \multicolumn{2}{r}{ $\mathrm{BG_{lep}}$  } & requirement & $\epsilon_q$ & $\mathrm{BG_{had}}$ \\ 
\hline
                           &            & 100 &         100 & \multicolumn{2}{r}{  142 M } &          &  100 &    142 M  \\
chg. PFOs                  & $4 \to 7$     &  91 &          93 & \multicolumn{2}{r}{ 10.1 M } &  $\geq 8$ &   98 &   95.7 M  \\ 
$Z\to ll$ candidate        &  $\geq 1$  &  88 &          90 & \multicolumn{2}{r}{ 1.03 M } &          &      & \\
isolated prongs            &            &     &             & \multicolumn{2}{r}{        } & $\geq 2$ &   91 &   45.8 M  \\
opp. chgd. prongs          &            &  84 &          87 & \multicolumn{2}{r}{  903 k } &          &   84 &   33.5 M \\ 
min. prong score           &            &     &             & \multicolumn{2}{r}{        } & $>0.8$   &   77 &   14.5 M  \\
impact par. error          &  $<25\mu m$ &  76 &          79 & \multicolumn{2}{r}{  491 k } &$<25\mu m$ & 74 &   13.2 M \\ 
extra cone energy          &            &  72 &          75 & \multicolumn{2}{r}{  438 k } &          &      & \\ 
$m_Z$                      &            &     &             & \multicolumn{2}{r}{        } & $60 \to 160$ &   72 &   5.58 M  \\
$m_\mathrm{recoil}$         &            &     &             & \multicolumn{2}{r}{        }& $50 \to 160$ &  71 &   4.90 M  \\
$\tau$ decay mode          &            &  63 &          65 & \multicolumn{2}{r}{  236 k } &           &   64 &   1.99 M \\ 
\hline
\hline
full selection & & \multicolumn{2}{c}{$Z \to ee$}& \multicolumn{2}{c}{$Z \to \mu\mu$} & & \multicolumn{2}{c}{$Z \to qq$} \\
event property & requirement & $\epsilon_e$   &  $\mathrm{BG}_e$           & $\epsilon_\mu$    & $\mathrm{BG}_\mu$ &  requirement & $\epsilon_q$  & $\mathrm{BG}_q$  \\ 
\hline
good \tpm\ fit           &             & 57 &                 112 k &       59 &       99.5 k &            & 58 &  1.64 M  \\
 $m_{\tau\tau}$          &  $100 \to 140$ &  46 &                  618 &       52 &         366  &  $100 \to 140$ &  42 &   42.9 k   \\
 event $p_T$             &     $<5$    & 43 &                   309 &       50 &         268  &     $<20$    & 42  &     30.9 k  \\
$m_\mathrm{recoil}$      &   $> 120$    &  42 &                 252 &       50 &         162  &   $> 100$    &   41 &    22.8 k  \\
$m_Z$                    &   $80 \to 105$   &  41 &                 186 &       49 &         136  &    $80 \to 115$  &  38  &  6.34 k  \\
$|\cos \theta_Z|$        &   $< 0.96$    &  40 &                 168 &       47 &         124  &   $< 0.96$   &  37  &     5.64 k  \\
 event $p_z$             &   $<40$       &  40 &                 144 &       47 &         105  &   $<40$      &  37  &      4.69 k  \\
 $|\cos \theta_{P}|_\mathrm{min}$ &   $< 0.95$   &   40 &                 140 &       47 &         102  &   $< 0.95$   &  37  & 4.69 k  \\
%
%
%good \tpm\ fit           &             & 57 &                 112 k &       59 &       99.5 k &            & 58 &    1.64 M  \\ 
% $m_{\tau\tau}$          &  $100 \to 140$ &  46 &                  618 &       52 &         366  &  $100 \to 140$ &  42 &     43.5 k   \\ 
% event $p_T$             &     $<5$    & 43 &                   309 &       50 &         268  &     $<20$    & 42  &      31.6 k  \\ 
%$m_\mathrm{recoil}$      &   $> 120$    &  42 &                 252 &       50 &         162  &   $> 100$    &   41 &      23.5 k  \\ 
%$m_Z$                    &   $80 \to 105$   &  41 &                 186 &       49 &         136  &    $80 \to 115$  &  38  &       6.93 k  \\ 
%$|\cos \theta_Z|$        &   $< 0.96$    &  40 &                 168 &       47 &         124  &   $< 0.96$   &  37  &       6.22 k  \\ 
% event $p_z$             &   $<40$       &  40 &                 144 &       47 &         105  &   $<40$      &  37  &       5.26 k  \\ 
% $|\cos \theta_{P}|_\mathrm{min}$ &   $< 0.95$   &   40 &                 140 &       47 &         102  &   $< 0.95$   &  37  &       5.26 k  \\ 
\hline
Sample purity (\%) &     &  \multicolumn{2}{c}{19} &  \multicolumn{2}{c}{26}  &  &   \multicolumn{2}{c}{11} \\ 
\end{tabular}
\end{ruledtabular}
\end{table*}%%%End of the table                          

\begin{figure}
\centering
\includegraphics[width=0.49\textwidth]{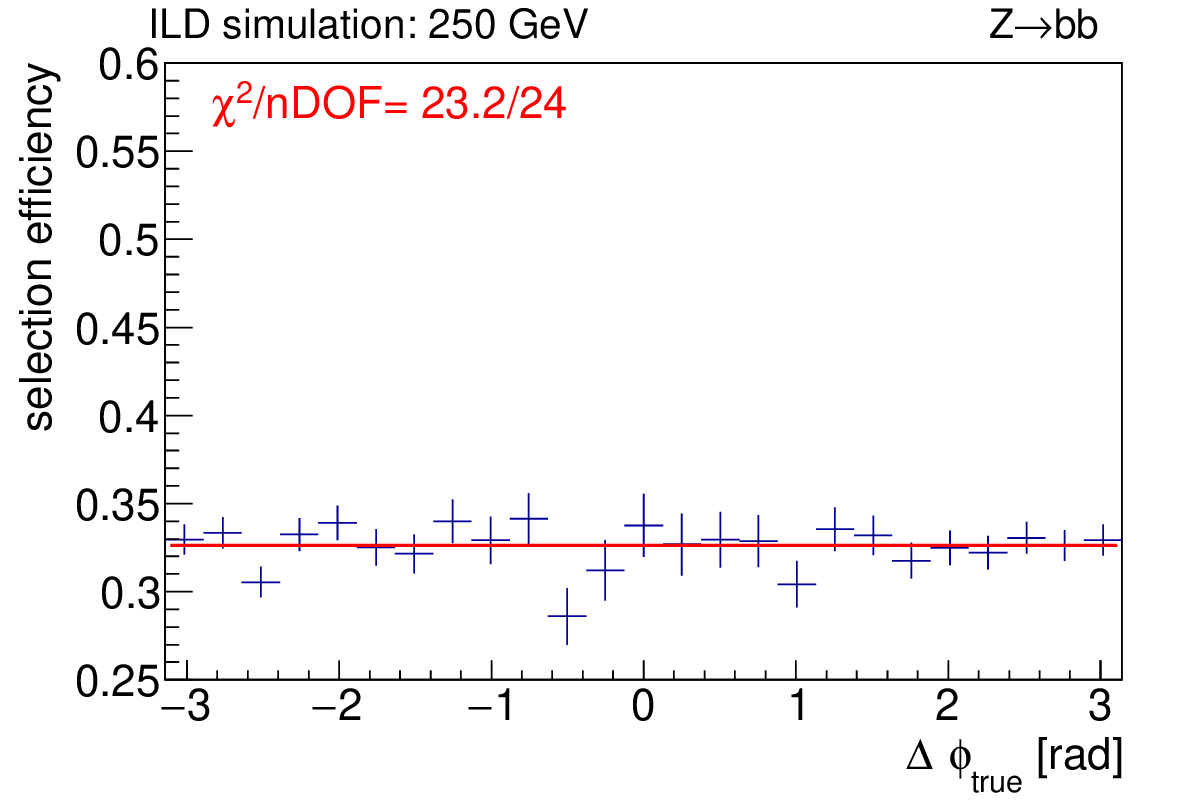} \\
\hspace{0.5cm}\\
\includegraphics[width=0.49\textwidth]{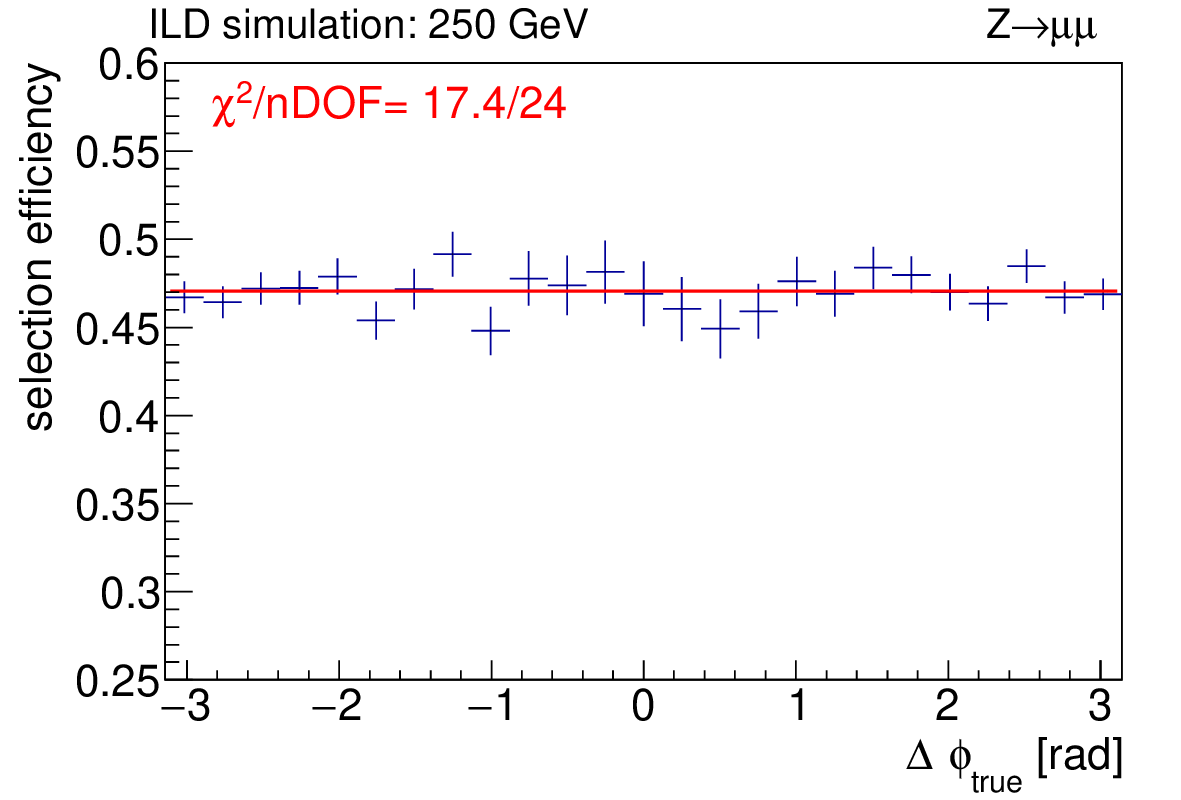} 
\caption{Signal reconstruction and selection efficiency as a function of the true value of 
\Dphi, in events with $Z \to bb$ and $Z \to \mu \mu$ decays. 
Vertical error bars on data points are due to finite MC statistics. The quality of fits of the points to a constant value are shown.
}
\label{fig:eff_dphi}
\end{figure}

Figure~\ref{fig:eff_dphi} shows how the efficiency of the full event selection depends on the
true value of \Dphi. The measured efficiency distributions are consistent with a uniform value, and no significant dependence
on \Dphi\ is seen.

The contributions of various processes to the backgrounds remaining after the selection are shown in Table~\ref{tab:bgsources}.
These include $\mathrm{HZ, H\to\tau\tau}$ events in which the $\tau$ leptons have decayed in different modes, 
and other four-fermion and two-fermion processes. Some of these backgrounds depend on the beam polarization: in particular
for the hadronic channel, the sample purity in the \erpl\ polarization scenario is significantly higher than for \elpr, due to
the suppression of backgrounds from W-pair production.
It is therefore advantageous to separately analyse data from the different polarization scenarios before combining their results.

\begin{table}
\caption{
Breakdown of remaining signal and background events after the three full selections.
Each line excludes processes which contribute to an earlier one.
Event numbers are scaled to the 2~\invab\ of 250~GeV data of the ``H20-staged'' running scenario, and are rounded to the nearest integer.
}
\label{tab:bgsources}
\centering
\begin{ruledtabular}
\begin{tabular}{lrrr}
process                                                           & $e$ & $\mu$ & $q$ \\
\hline
signal                                                            &  32        &   36           &  575 \\
other $\mathrm{f \overline{f} H, H \to \tpm}$ &  39        &   43           &  627 \\
other $\mathrm{f \overline{f} H}$             &   1        &    0           &   58 \\
other $\mathrm{f \overline{f} \tpm}$          &  32        &   24           &  766 \\
other $\mathrm{4 f}$                          &  51        &   35           & 2834 \\
$\mathrm{2 f}$                                &  18        &    0           &  403 \\
\end{tabular}
\end{ruledtabular}
\end{table}%%%End of the table                          

\subsection{Neural Networks}

\begin{figure*}
\centering
\includegraphics[width=0.49\textwidth]{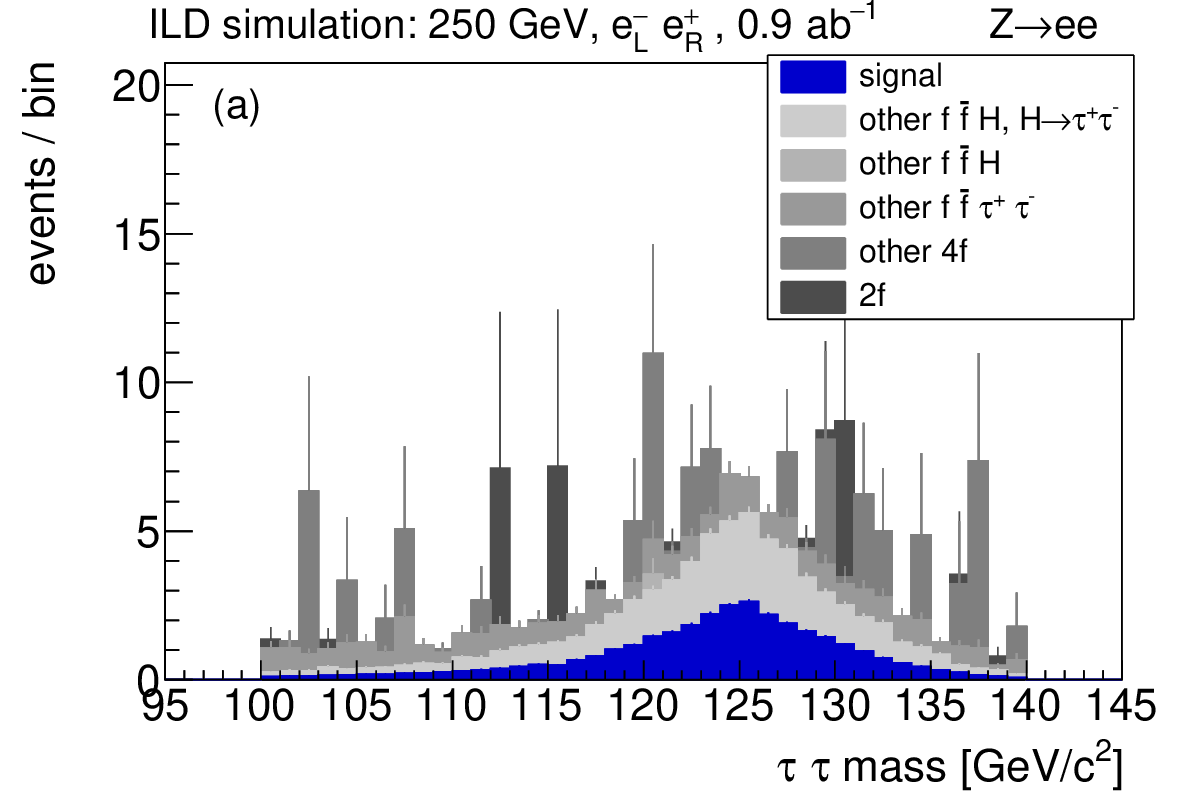}
\includegraphics[width=0.49\textwidth]{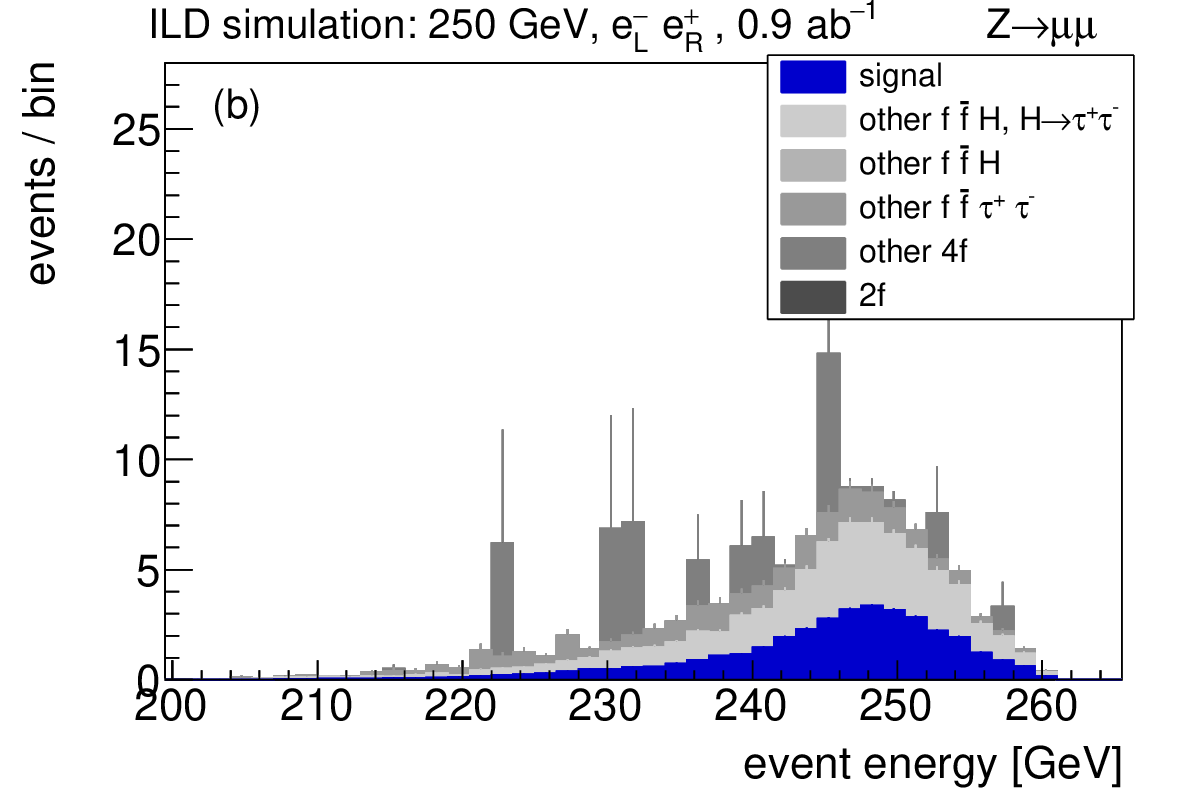}\\
\hspace{0.5cm}\\
\includegraphics[width=0.49\textwidth]{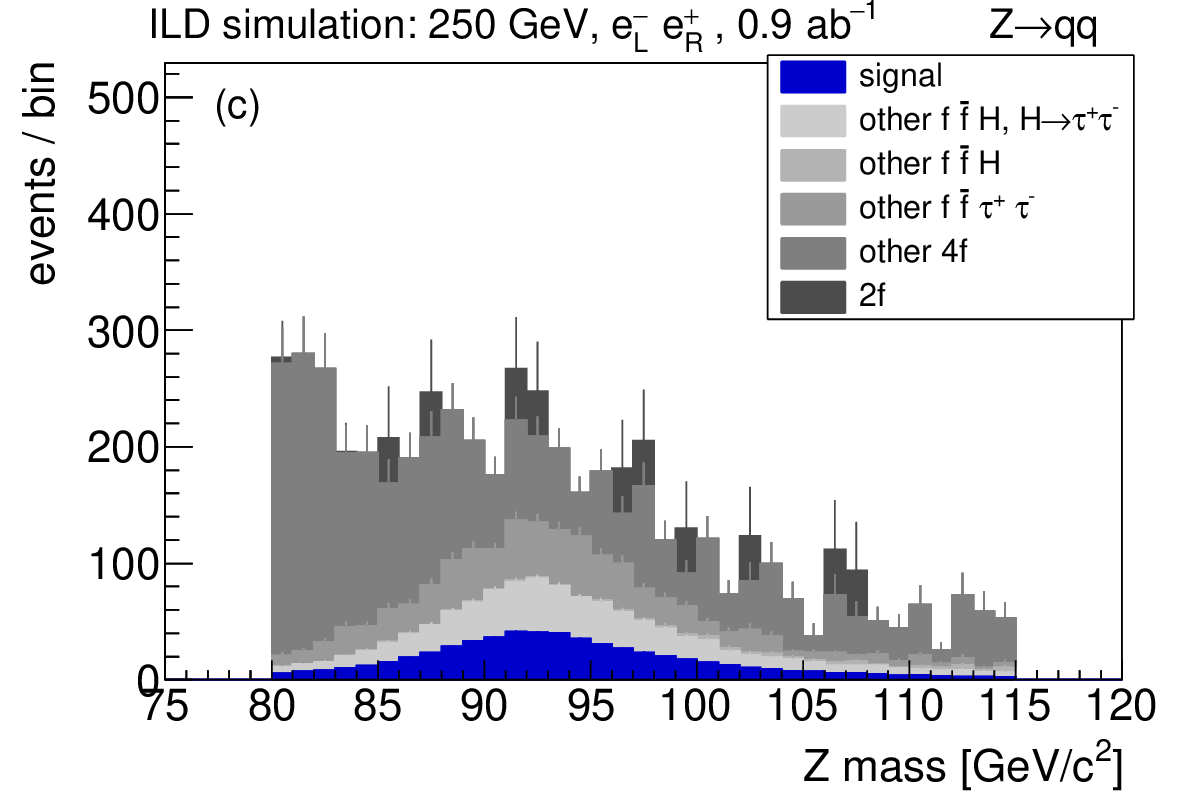}
\includegraphics[width=0.49\textwidth]{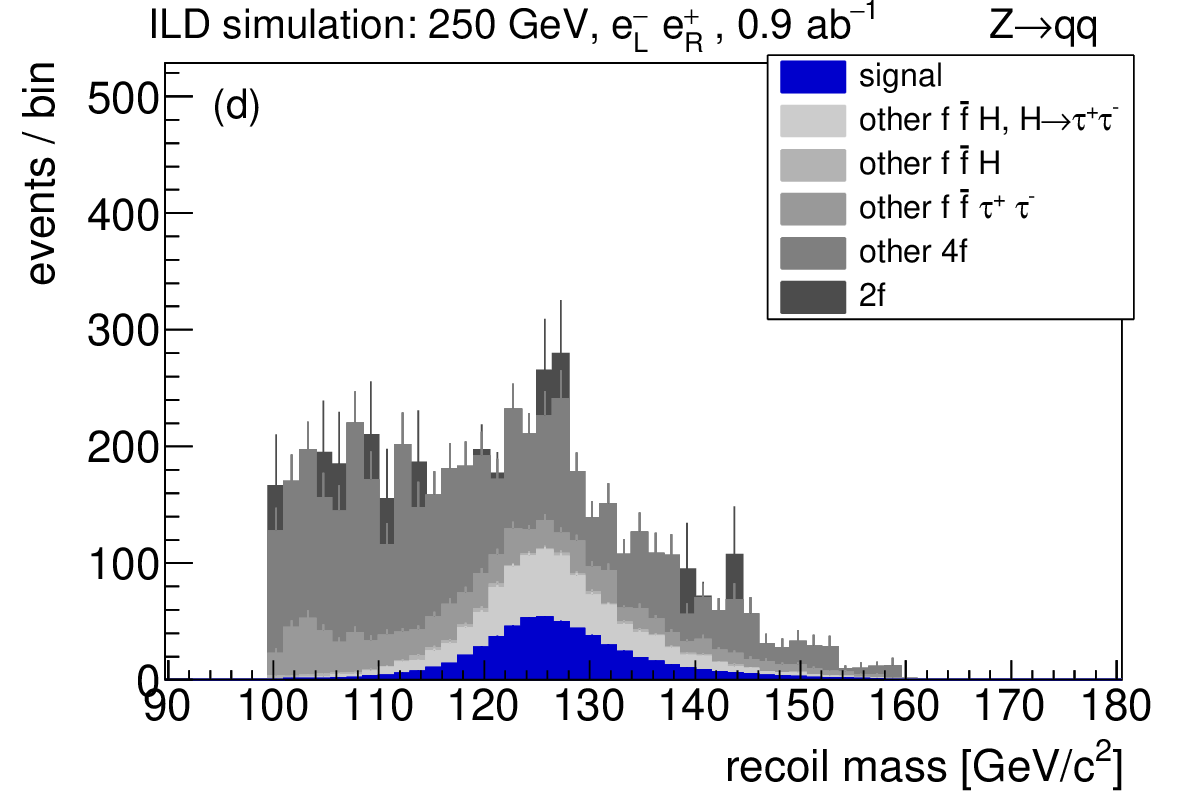}
\caption{Distribution of some input variables to the neural networks, for selected events. 
Distributions are normalized to 0.9~\invab\ of data in the \elpr\ beam polarization.
}
\label{fig:nntrainvars}
\end{figure*}

\begin{figure*}
\centering
\includegraphics[width=0.49\textwidth]{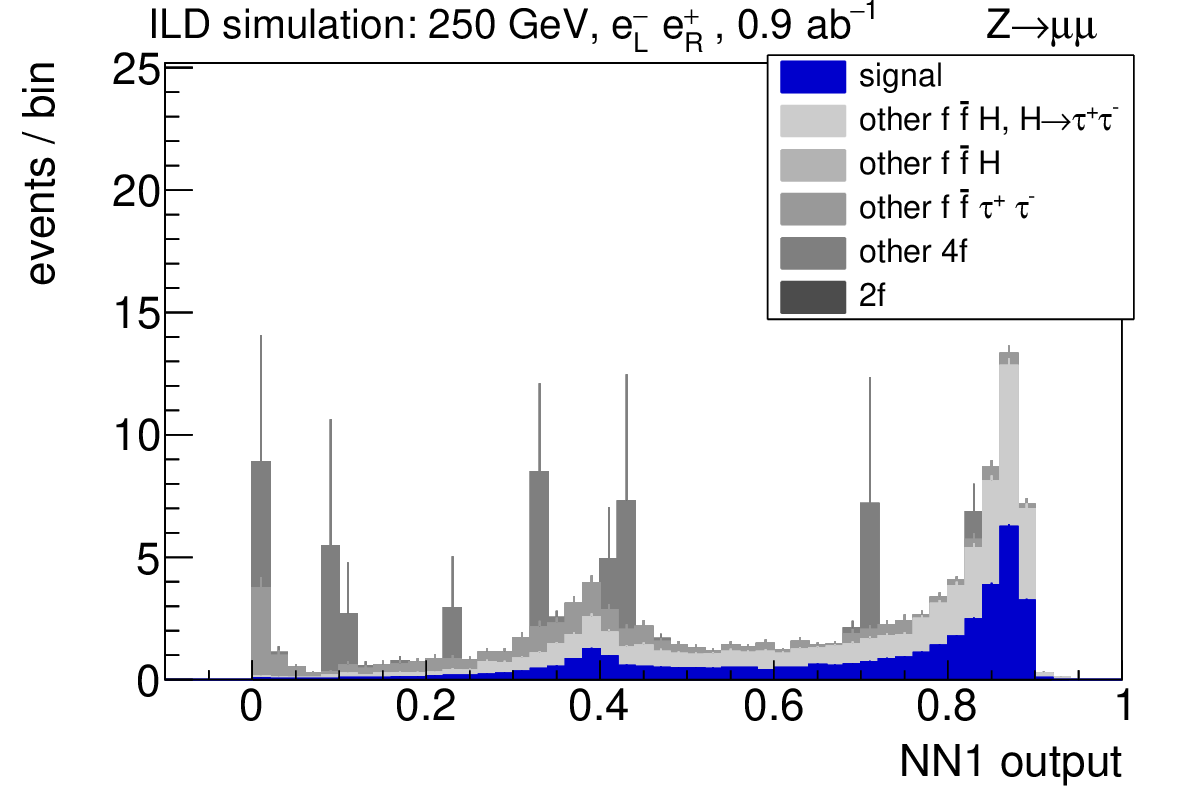}
\includegraphics[width=0.49\textwidth]{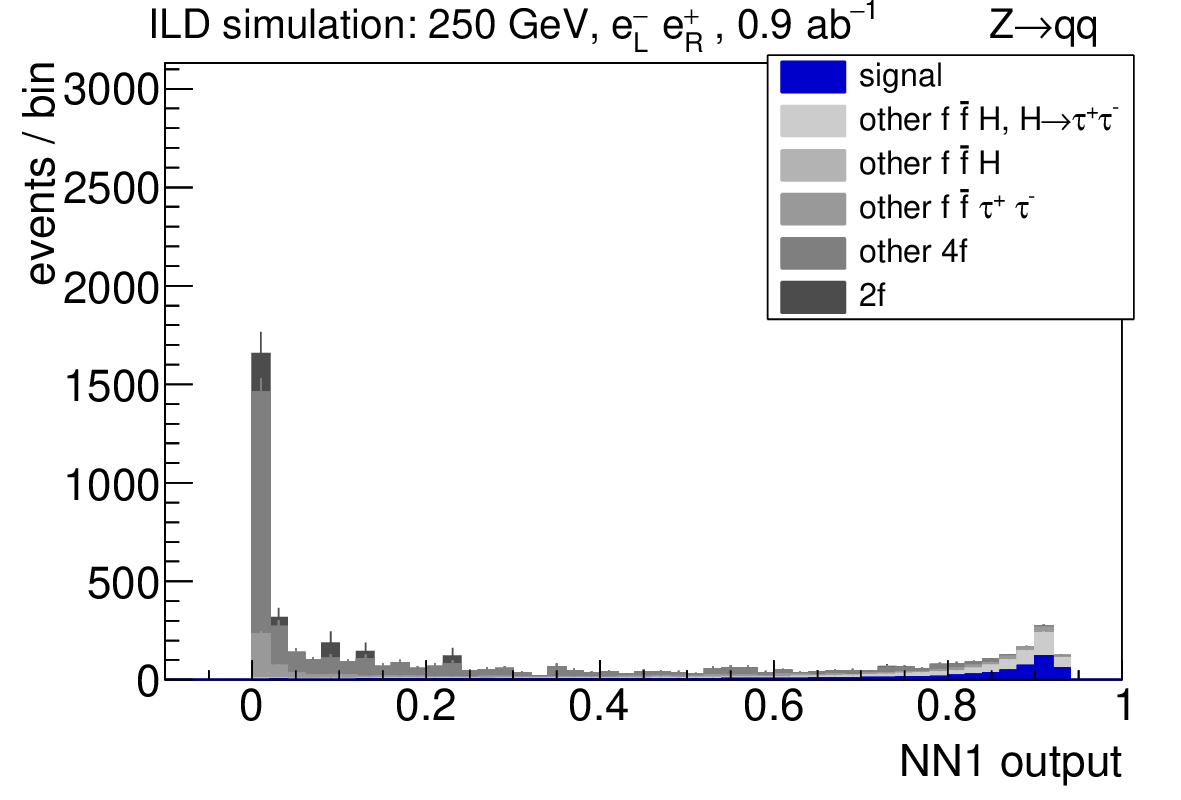}\\
\hspace{0.5cm}\\
\includegraphics[width=0.49\textwidth]{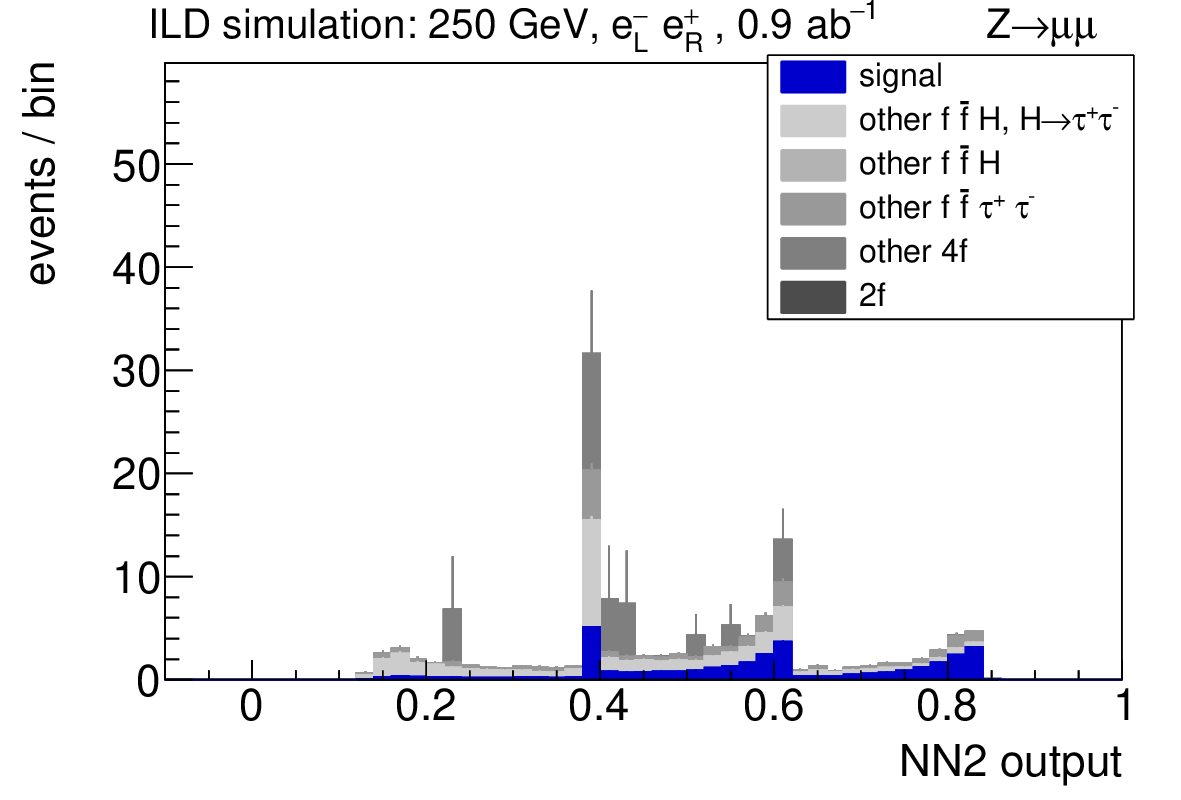}
\includegraphics[width=0.49\textwidth]{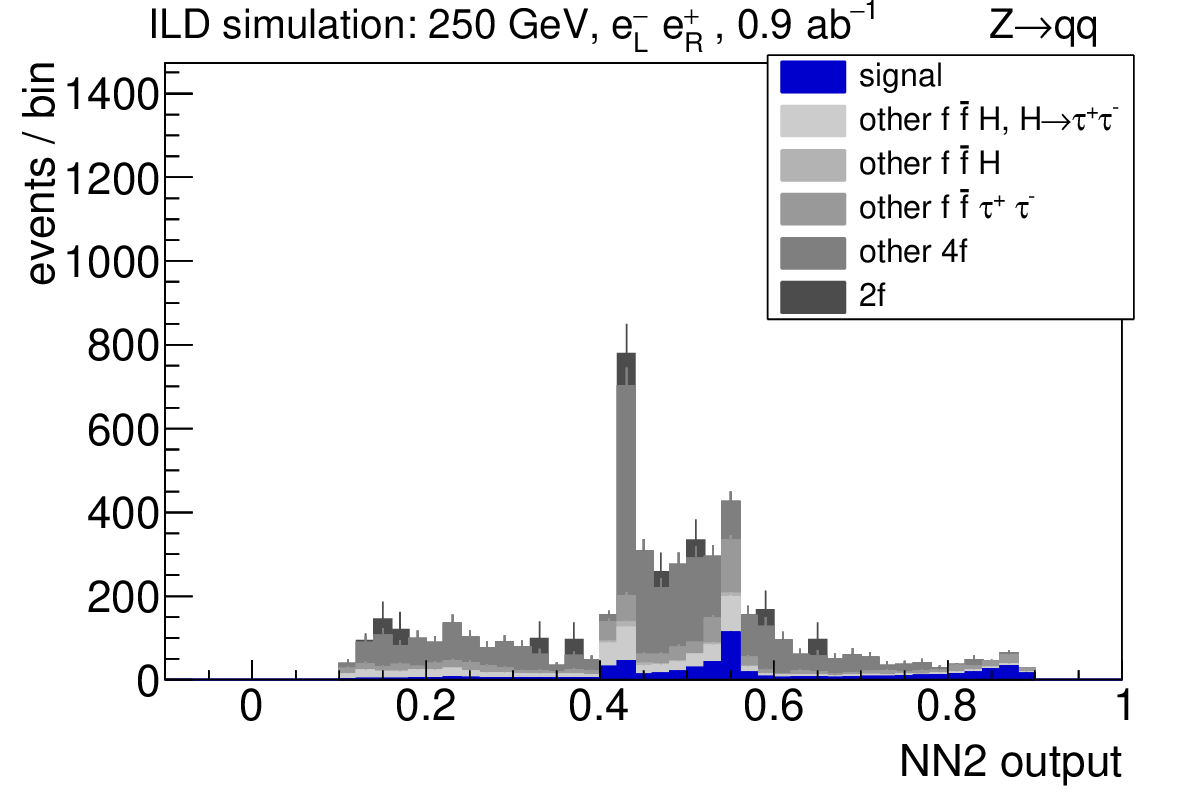}
\caption{Distributions of the two Neural Network outputs in the muon and hadronic selection channels.
The structure in the output of NN2 is due to the three different combinations of $\tau$ lepton decay modes.
Distributions are normalized to 0.9~\invab\ of data in the \elpr\ beam polarization.
}
\label{fig:NNout}
\end{figure*}

A pair of artificial Neural Networks (NN) were used to distinguish signal and background events which passed each channel's selection.
The first was trained to distinguish signal ZH events from major four-fermion backgrounds, and the second to distinguish ZH events 
with signal $\tau$ decay modes from those with other $\tau$ modes. 
Networks were separately trained for the ${\mathrm Z \to e^+ e^-}$, $\mu^+ \mu^-$, and hadronic channels.
The first network was trained using six input observables: the \tpm\ invariant mass, event energy, invariant mass of the recoiling system, 
the recoil mass, and, for each $\tau$, the sum of the energy of PFOs 
within $20^\circ$ of the $\tau$ prong but not assigned to the $\tau$ jet.
The second network was trained using four observables: the same $20^\circ$ energy sums and the visible mass of each tau jet.
In each NN, a single hidden layer was used, containing one less node than the number of input variables.
Distributions of some of these input variables are shown in Fig.~\ref{fig:nntrainvars}, and of the network outputs
in Fig.~\ref{fig:NNout}.

\begin{figure*}
\centering
\includegraphics[width=0.49\textwidth]{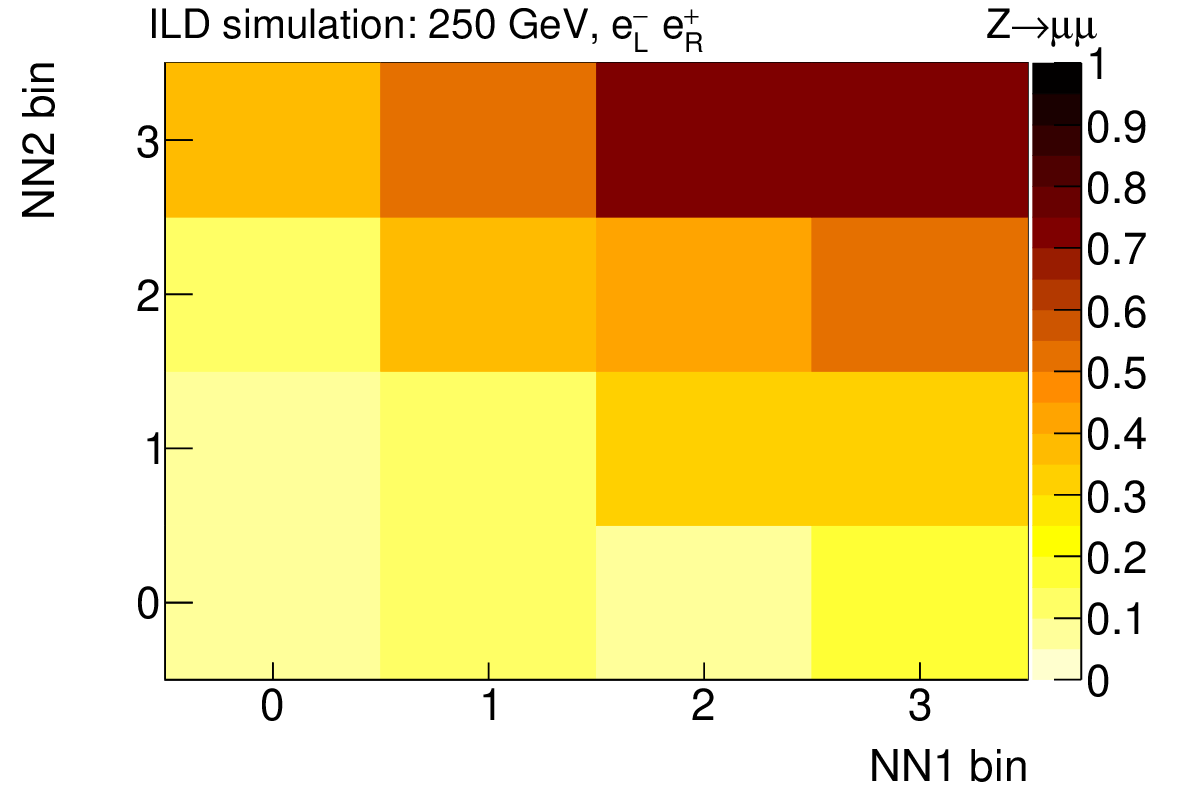}
\includegraphics[width=0.49\textwidth]{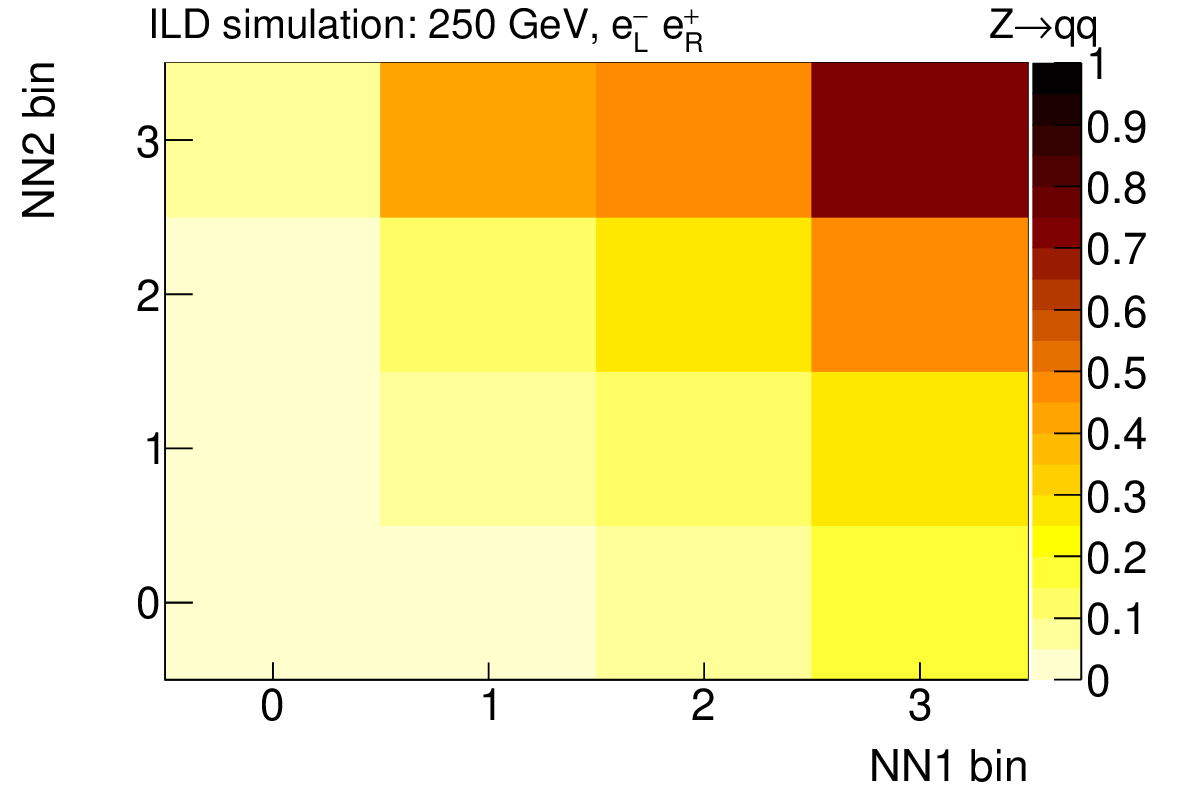}
\caption{
The color scale shows the signal purity in $4\times4$ bins of Neural Network outputs, in the muon and hadronic 
selection channels.
}
\label{fig:qqNN_purity}
\end{figure*}

Within each selection channel, events were split into $4 \times 4$ classes according to the outputs of the two NNs.
These classes have rather different signal purities, and therefore different sensitivities to the CP effects being measured,
as shown in Fig.~\ref{fig:qqNN_purity}.

\subsection{CP observables}

The polarimeter vector of each $\tau$ lepton was reconstructed in its rest frame according to its assigned decay mode, using Eqn.~\ref{eqn:polar}.
In this frame, the angles $\theta$ and $\phi$ of the $\tau^\pm$ polarimeters with respect to $(\pm)$ the momentum of the \tpm\ system were used as sensitive observables.
As discussed in Section~\ref{sec:cpang},
the observable $\Dphi = \phi^+ - \phi^-$ is sensitive to the CP mixing angle \psicp, while the sensitivity of a particular event depends on the value of the contrast function $\cthth$.

\begin{figure*}
\centering
\includegraphics[width=0.49\textwidth]{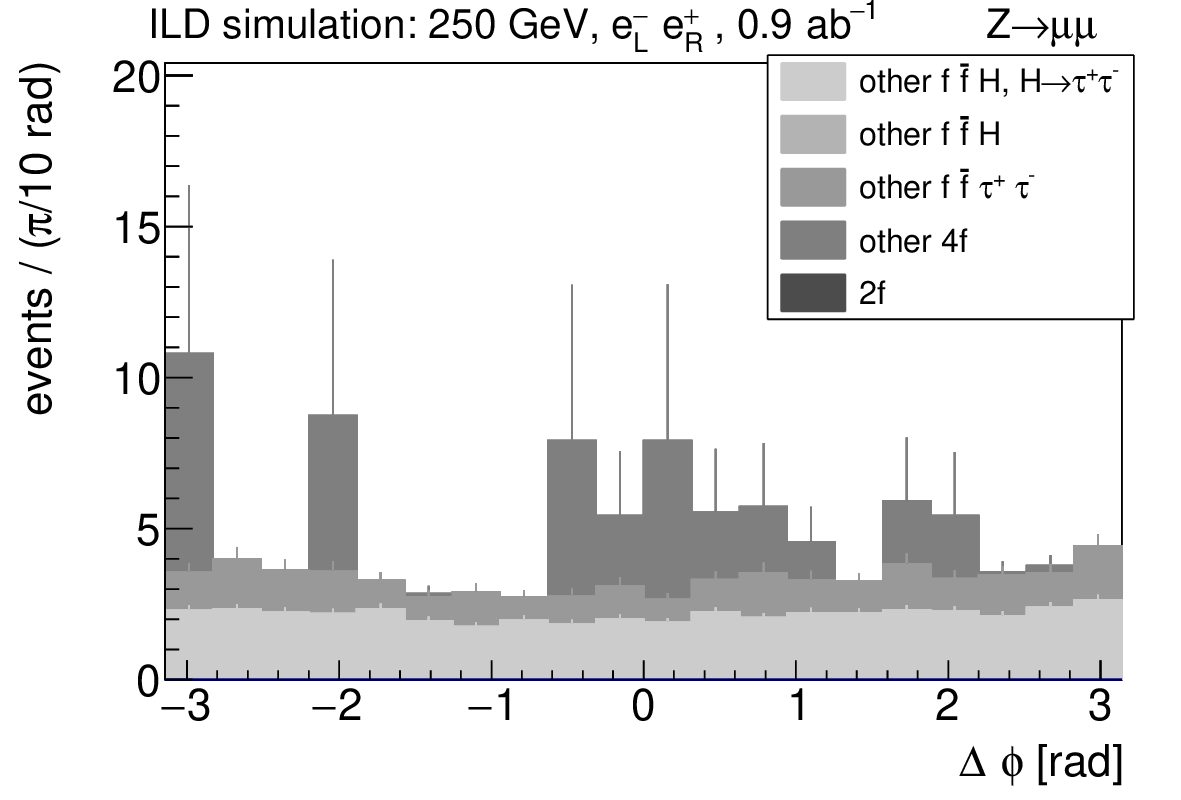}
\includegraphics[width=0.49\textwidth]{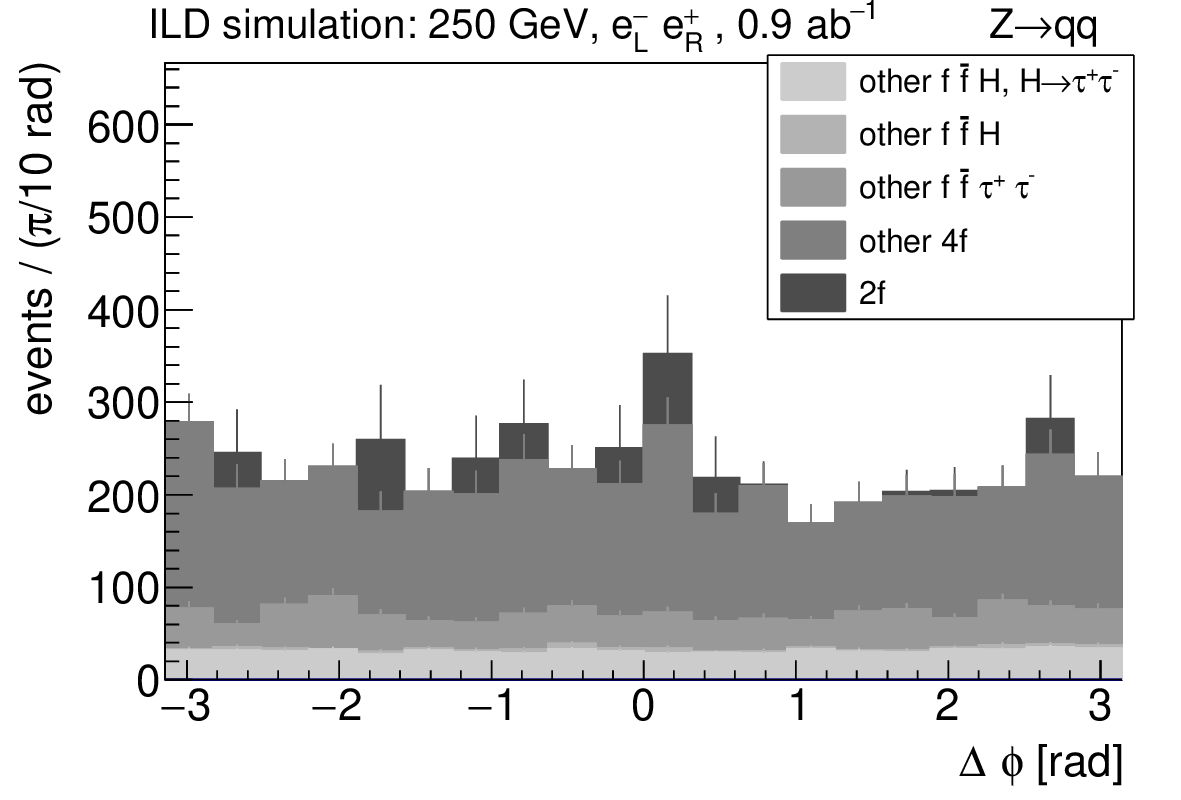}
\caption{\Dphi\ distributions of background events selected in the muon and hadronic selection channels.
Distributions are normalized to 0.9~\invab\ of data in the \elpr\ beam polarization.
Error bars reflect uncertainties due to limited MC simulation statistics.
}
\label{fig:dphi_nosig}
\end{figure*}

Distributions of selected background events in \Dphi\ are shown in Fig.~\ref{fig:dphi_nosig}.
No significant modulation is seen, given the uncertainties due to finite simulation statistics. In particular,
the $\mathrm{f\overline{f}H}$ contribution with H decays to $\tau$ leptons but $\tau$ decays to other final states than those
used in this analysis (``other $\mathrm{ f \overline{f} H, H \to \tau^+ \tau^-}$'') is flat. Even though the $\tau$ lepton pairs have been
decayed including the approriate spin correlations, the use of inappropriate polarimeters washes out the correlation.

\begin{figure}
\centering
\includegraphics[width=0.49\textwidth]{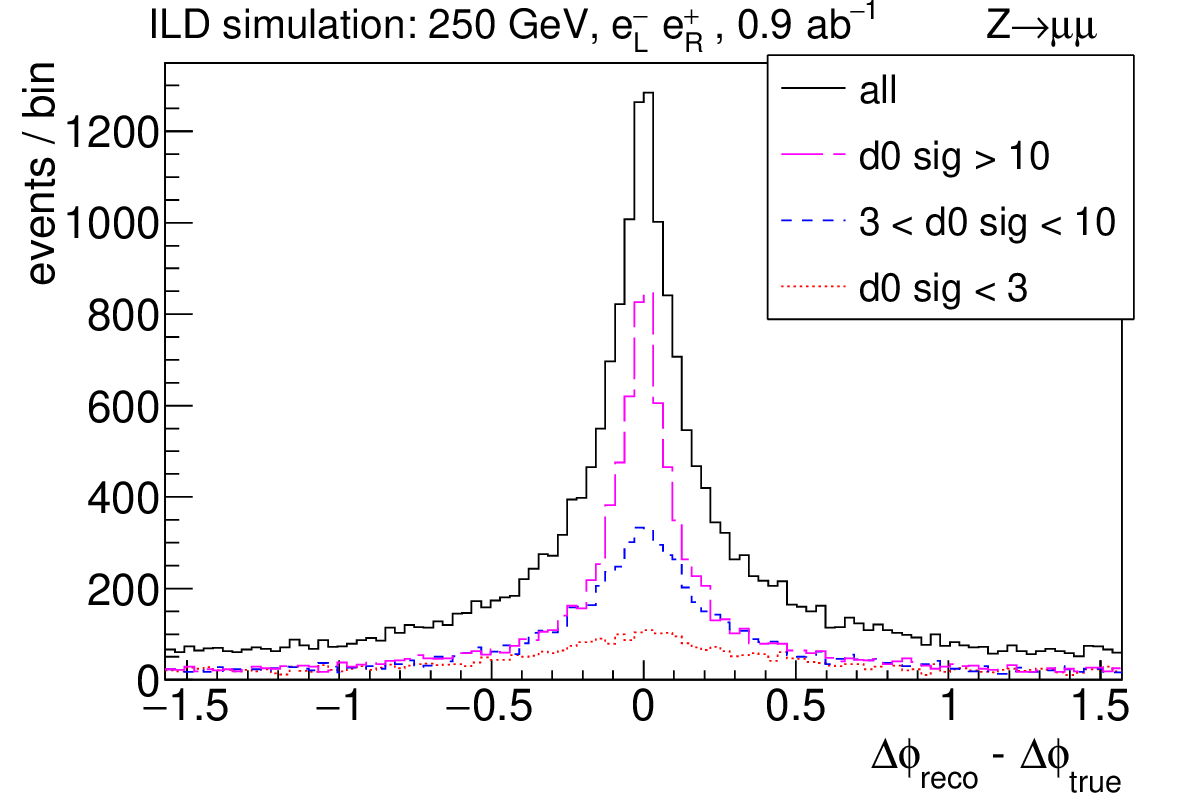}
\caption{
Difference between reconstructed and true \Dphi\ values, in different ranges of \dsm.
}
\label{fig:dphi_res}
\end{figure}

\begin{figure}
\centering
\includegraphics[width=0.49\textwidth]{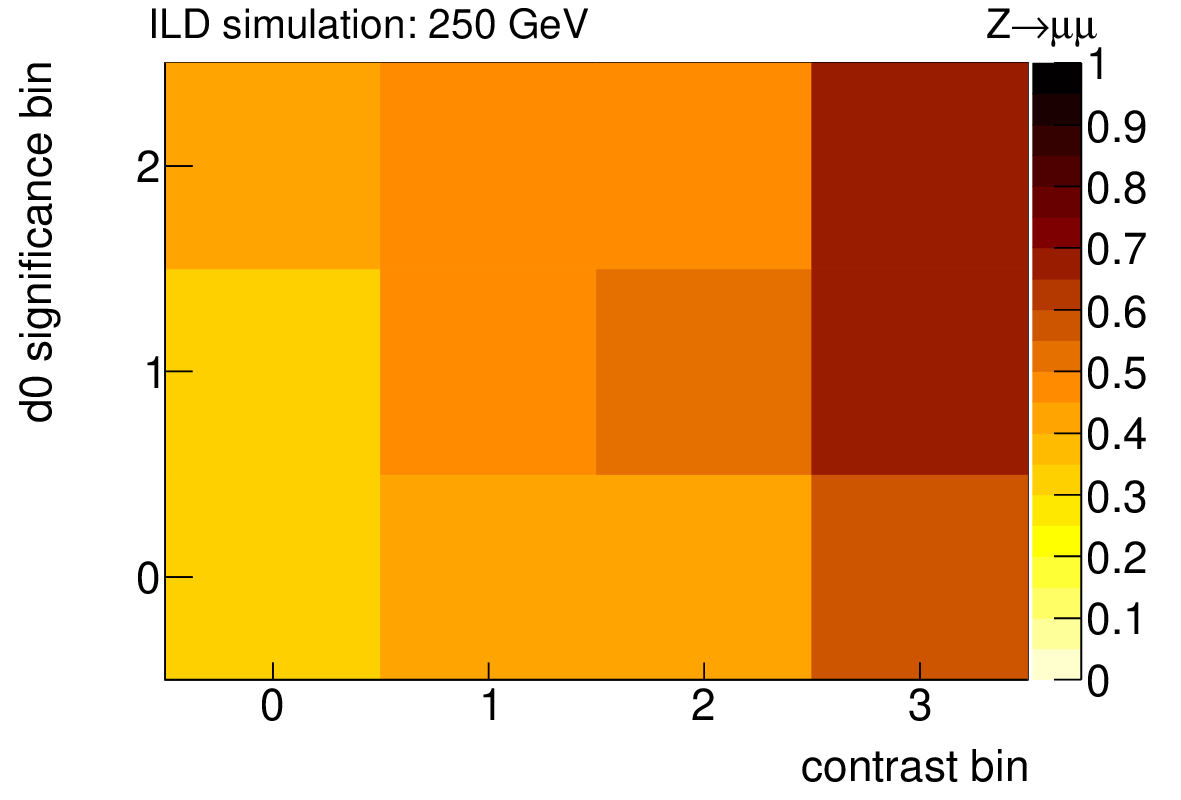}
\caption{
The color scale shows the modulation amplitude of the reconstructed \Dphi\ distribution of signal events in different bins of contrast and \dsm.
}
\label{fig:dphi_mod}
\end{figure}

The precision with which \Dphi\ is measured in a particular event depends on the quality of the $\tau$ lepton reconstruction.
The most delicate input to this reconstruction is the trajectory of the $\tau$ decay prong, which is used to define the plane within which the $\tau$ momentum lies. 
If a $\tau$ decay prong is well separated from the PV, the $\tau$ decay plane can be more precisely reconstructed than if the
prong has only a small, or a poorly measured, displacement. This could occur, for example, if the $\tau$ has only a short lifetime
or if the prong's impact parameter was not well measured. 
A suitable parameter to quantify this effect is the significance of the prong's impact parameter from the PV, $d_\mathrm{sig} \equiv |d_0| / \sigma_{d0}$,
where $d_0$ is the track's impact parameter and $\sigma_{d0}$ the uncertainty on its measurement.
Figure~\ref{fig:dphi_res} shows how the difference between the reconstructed and true values of \Dphi\ depends on \dsm, 
the smaller of the two prongs' $d_\mathrm{sig}$.
For $\dsm < 3$, the resolution on \Dphi\ is very poor, while it is much better (of order $100$~mrad) for $\dsm > 10$.
The distribution is symmetrical, demonstrating that the reconstruction is unbiased.
Events were split into three classes according to $\dsm$, with boundaries at 3 and 10, 
and into four classes according to the reconstructed value of the contrast function, with boundaries at 0.3, 0.6, and 0.9.
Figure~\ref{fig:dphi_mod} shows how the reconstructed modulation amplitude of signal events' \Dphi\ distributions varies in these $3 \times 4$ bins.

%As explained above, preselected events were classified according to the outputs of two Neural Networks, the smaller 
%of the two prongs' impact parameter significance $|d_0| / \sigma_{d0}$, and the reconstructed value of the contrast function. 
%Each of the resulting large number of sub-classes has a different sensitivity to \psicp, 
%determined by the modulation amplitude of the total (signal plus background) \Dphi\ distribution.
%For each selection channel, these sub-classes were distributed into four groups according to this total modulation amplitude.

As explained above, selected events in each channel were assigned to one of $4 \times 4 \times 3 \times 4$ categories, depending respectively on the
outputs of the two neural networks, the impact parameter significance, and the value of the contrast function.
In each category, the total (signal + background) \Dphi\ distribution was fitted with a function of form
$y = a ( 1 - C \cos \Dphi ) $. 
The expected relative amplitude of the modulation, $C$, was extracted for each category.
Categories with similar values of $C$, and therefore similar per-event sensitivity to CP effects, were combined into four larger groups.
If the relative amplitude $C$ of a category was larger than 0.3 it was assigned to group A; if between 0.2 and 0.3 to group B;
between 0.1 and 0.2 to group C; and the remainder to group D.

\begin{figure*}
\centering
\includegraphics[width=0.49\textwidth]{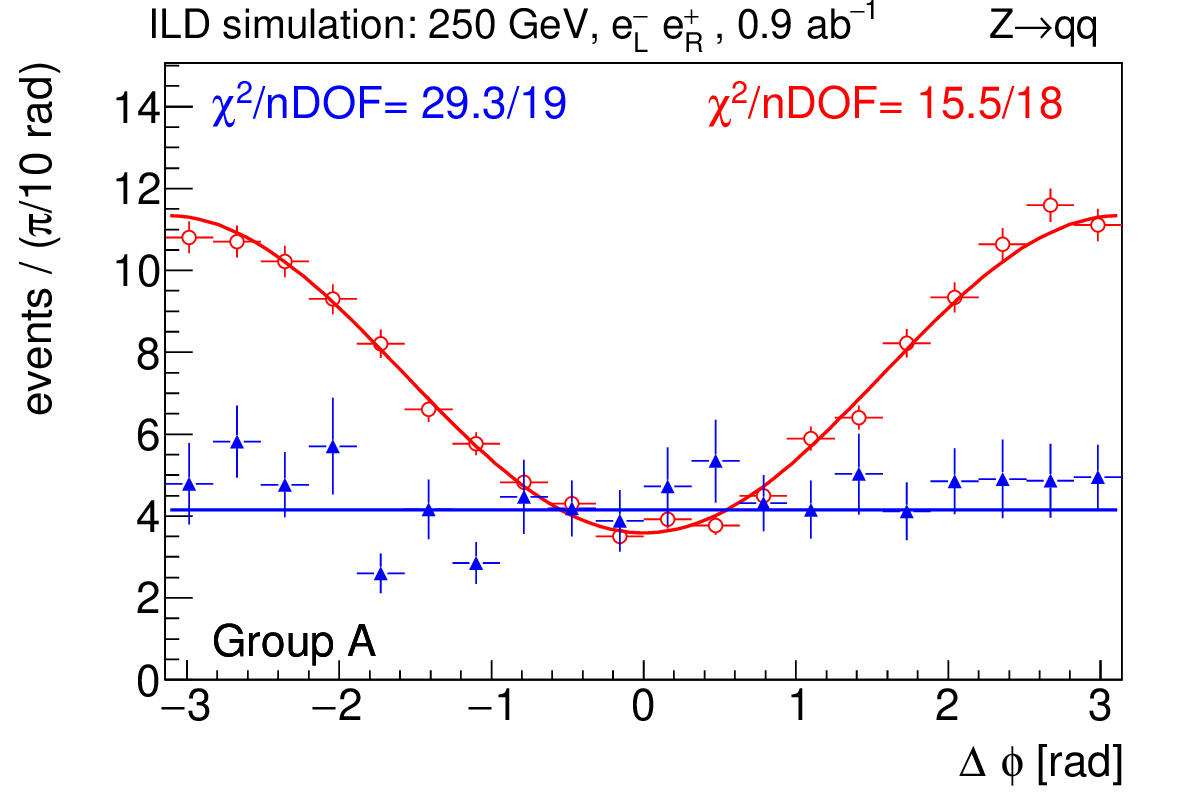}
\includegraphics[width=0.49\textwidth]{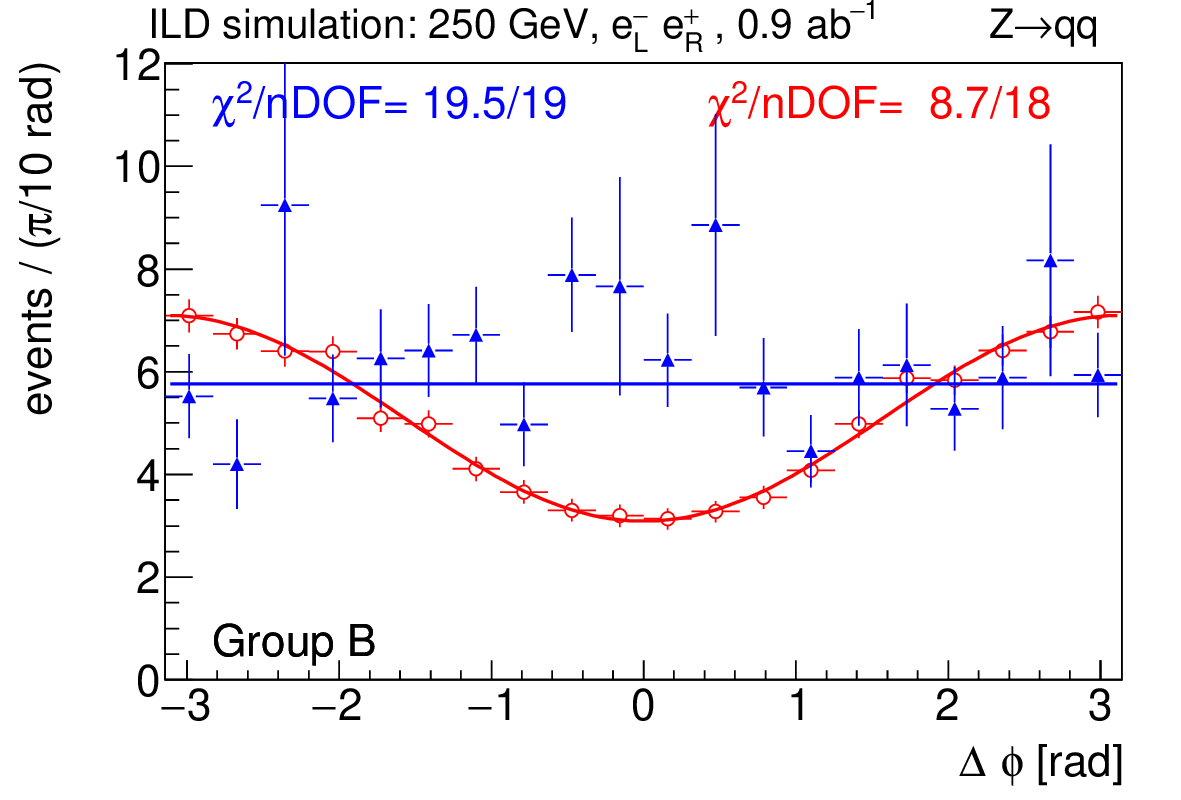}\\
\hspace{0.5cm}\\
\includegraphics[width=0.49\textwidth]{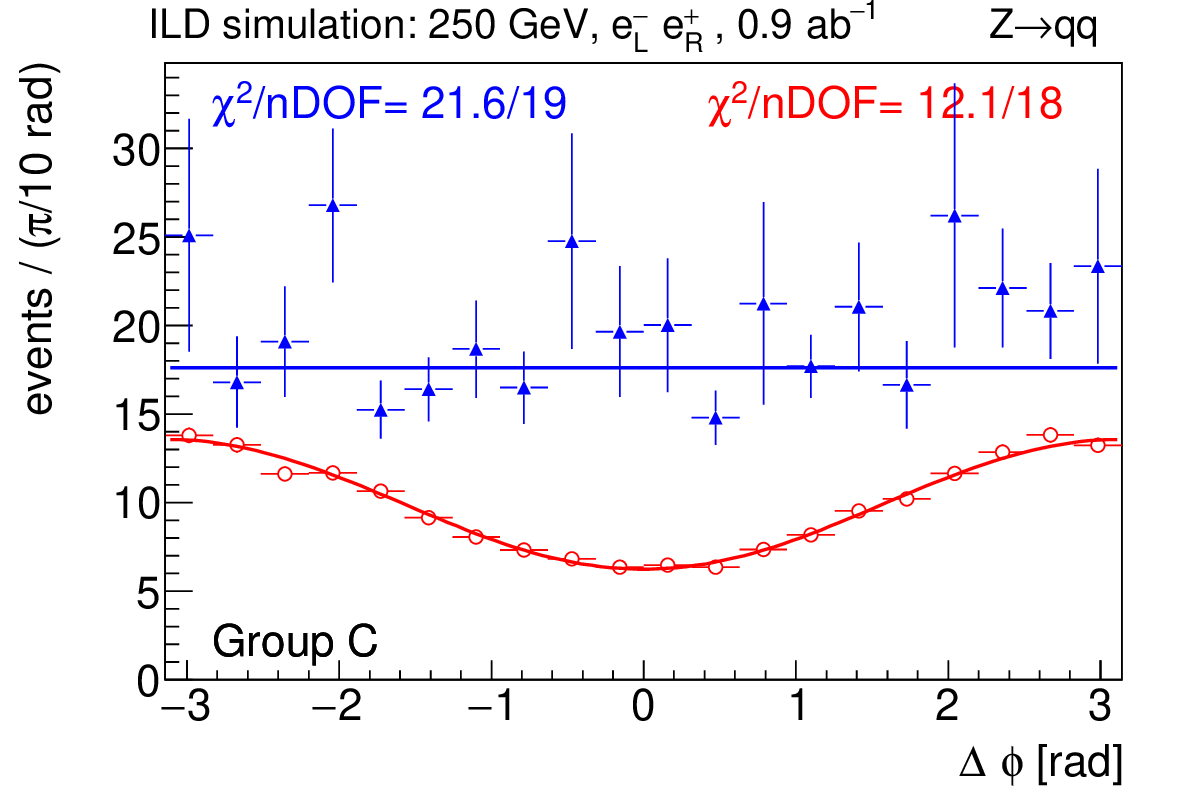}
\includegraphics[width=0.49\textwidth]{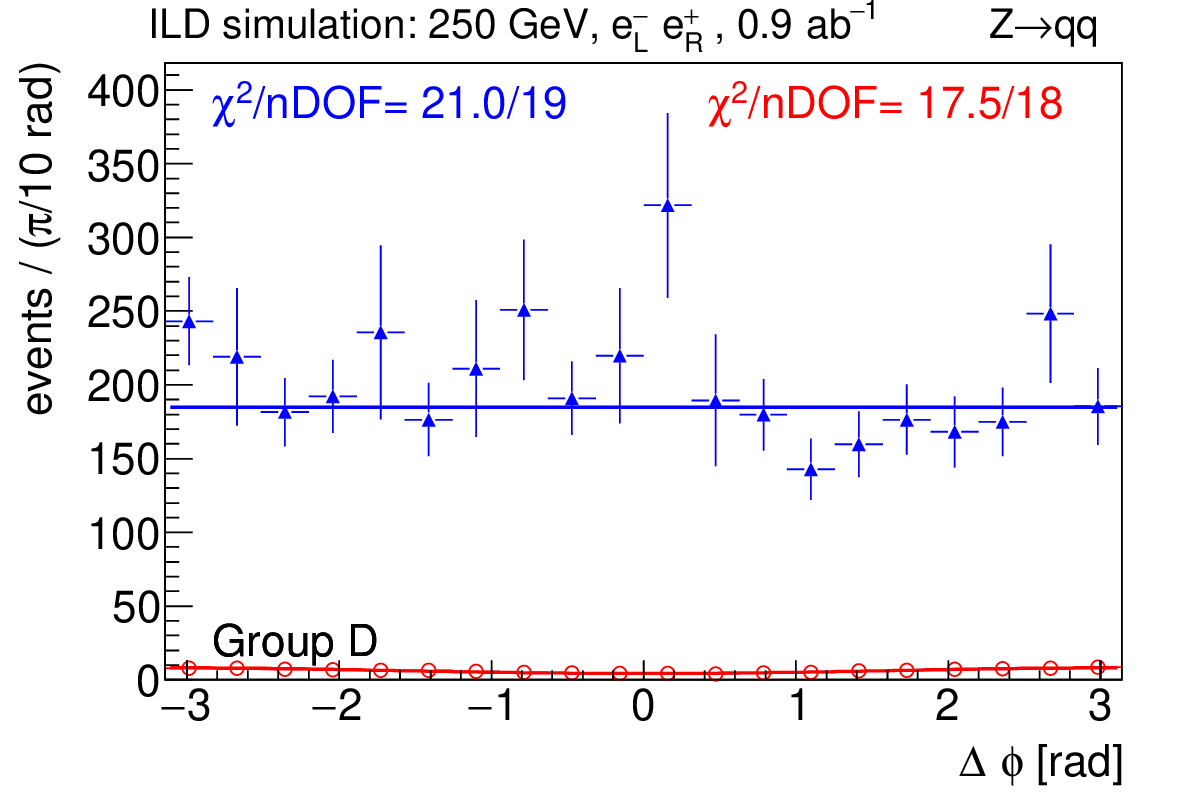}
\caption{Distributions of reconstructed \Dphi\ in the four modulation amplitude groups,
for events selected in the hadronic channel. The distribution of signal events is shown as open red circles, 
together with the result of the fit described in the text.
The total background distribution is shown as filled blue triangles, together with a fit to a constant value.
The $\chi^2$/nDOF of these fits is also reported.
The error bars reflect uncertainties due to finite statistics of the simulation samples. 
Distributions are normalized to 0.9~\invab\ of data in the \elpr\ beam polarization.
Signal samples were generated with $\psicp=0$ (i.e. the SM).
}
\label{fig:dphi}
\end{figure*}

Distributions of \Dphi\ for signal and background in the four groups of the hadronic channel are shown in Fig.~\ref{fig:dphi}.
The backgrounds are consistent with a flat distribution, and do not show the modulated shape of the signal. 
The signal modulation amplitude and signal-to-background ratio vary considerably among the four groups. 
The sample of events in the group least sensitive to CP (labeled as ``Group D'') has negligible sensitivity to CP, and was 
not used in the CP measurement.

\section{Sensitivity at ILC}
\label{sec:results}

\begin{figure}
\centering
\includegraphics[width=0.49\textwidth]{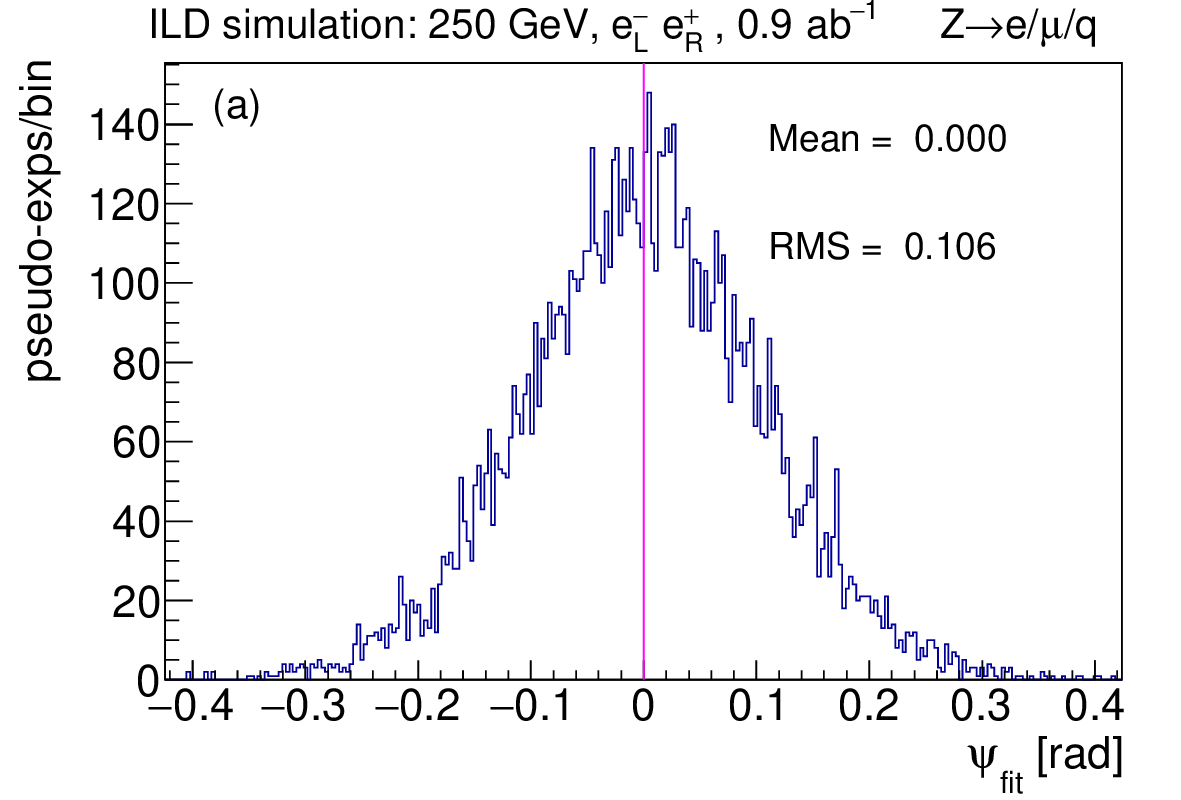} \\
\hspace{0.5cm}\\
\includegraphics[width=0.49\textwidth]{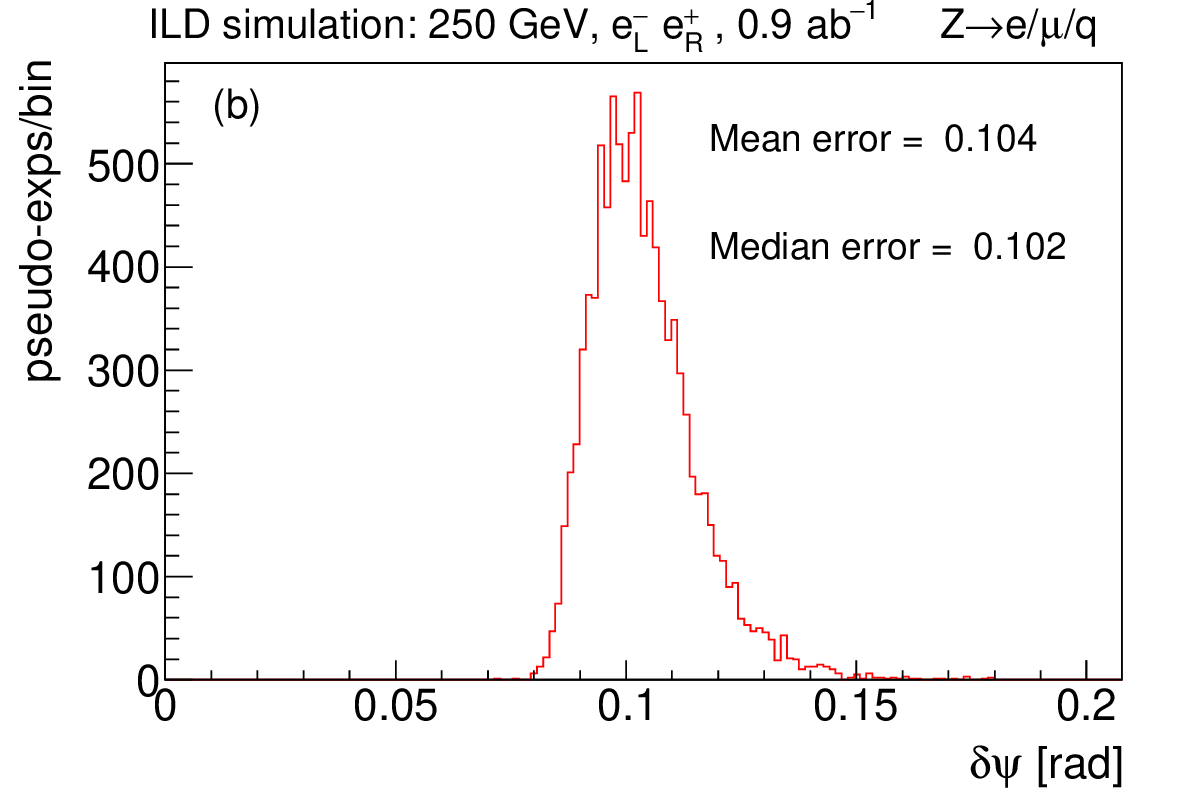} \\
\hspace{0.5cm}\\
\includegraphics[width=0.49\textwidth]{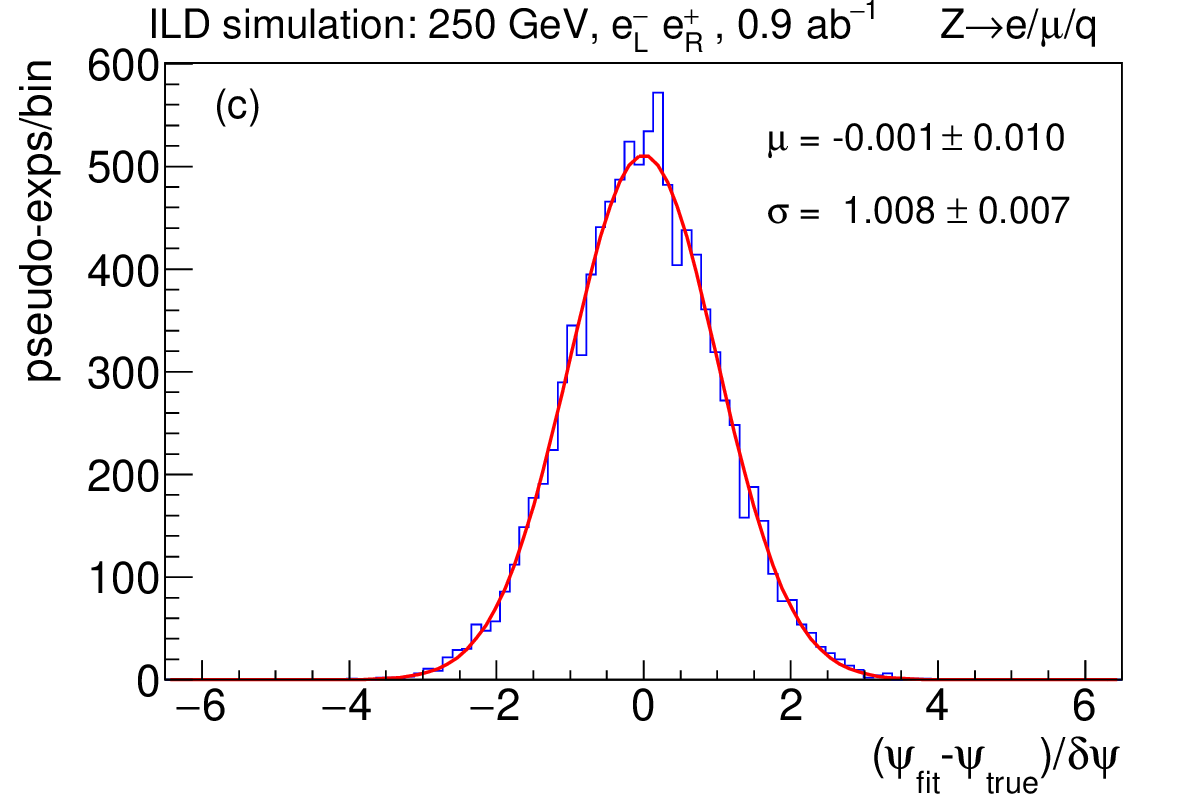}
\caption{Results of 10~k pseudo-experiments, for a combined fit to all three selection channels,
for 0.9~\invab\ of data in the \elpr\ beam polarization:
(a) the extracted value of \psicp\ -- the vertical line shows the true input value of 0; 
(b) $\delta \psi$, the extracted uncertainty on \psicp; and (c) the pull distribution,
overlaid with a fit to a Gaussian function.}
\label{fig:toyresults}
\end{figure}

Distributions such as those in Fig.~\ref{fig:dphi} were used to estimate the precision with which
$\psi_\mathrm{CP}$ can be measured by means of pseudo-experiments.
The distribution of signal events was fitted to a function of form $f(\Dphi) \propto 1 - C \cos(\Dphi - 2 \psicp)$
with the CP mixing angle \psicp\ fixed to the input (SM) value of 0.
The background was assumed to be uniformly distributed in \Dphi. 
The three \Dphi\ distributions (groups A--C) considered for each selection channel and polarization scenario were used to run a series of pseudo-experiments.
The mean number of events expected in each sub-sample was calculated according to the assumed integrated luminosity. 
In each sub-sample, a number of events, Poisson distributed around this mean, was distributed according to the expected distribution
in \Dphi. An unbinned maximum likelihood fit to these events was used to extract an estimate of \psicp, 
simultaneously fitting all sub-samples of a given polarization scenario. 
The expected contrast ($C$) in each sub-sample, determined from the fit to the pseudo-experiments' parent distributions, 
was treated as a fixed parameter, leaving \psicp, the phase of the distribution, 
as the only free parameter. The results of such pseudo-experiments are shown in Fig.~\ref{fig:toyresults}, in which the 
distribution of the extracted phase, its uncertainty, and their ratio are shown.
The extracted value of the CP phase shows no sign of bias, and the pull distribution demonstrates that uncertainties are well
estimated. 

\begin{table}
\caption{Estimated experimental precision $\delta\psicp$ on the CP phase in different scenarios.}
\label{tab:allresults}
\centering
\begin{ruledtabular}
\begin{tabular}{lrrlr}%%%The number of columns has to be defined here
$\int \mathcal{L}$
 & \multicolumn{2}{c}{beam pol.} & notes   & $\delta\psicp$ \\
$[ \invab ]$ &       $e^-$ & $e^+$ &  &         [mrad]     \\
\hline
1.0 & 0   & 0           & full analysis     & 116 \\
%\hline
1.0 &  0 & 0  & only $Z\to ee$       & 450 \\
1.0 &  0 & 0  & only $Z \to \mu\mu$ & 412 \\
1.0 &  0 & 0  & only $Z \to qq$     & 122 \\
%\hline
1.0 &  0 & 0  & only $(\pi\nu, \pi\nu  )$ & 387 \\
1.0 &  0 & 0  & only $(\pi\nu, \rho\nu )$ & 198 \\
1.0 &  0 & 0  & only $(\rho\nu, \rho\nu)$ & 166 \\
%\hline
1.0 & $-1.0$ & $+1.0$     & pure \elpr     & 97 \\
1.0 & $+1.0$ & $-1.0$     & pure \erpl     & 113 \\
%\hline
1.0 & 0   & 0           &  $\sigma_{ZH} + 20\%$    & 104 \\
1.0 & 0   & 0           &  $\sigma_{ZH} - 20\%$    & 133 \\
%1.0 & 0   & 0           &  $\sigma_{X H \to \tpm} + 20\%$    &  105 \\
%1.0 & 0   & 0           &  $\sigma_{X H \to \tpm} - 20\%$    &  134 \\
%\hline
1.0 &  0 & 0  & no bg.                         & 76 \\
1.0 &  0 & 0  & perf. pol.                     & 100 \\
1.0 &  0 & 0  & no bg., perf. pol./eff. & 25 \\
\hline
\hline
\multicolumn{5}{c} {H20-staged: 250 GeV, 2~\invab}  \\ % -- ignoring worst modulation bin
\hline
0.9 & $-0.8$ & $+0.3$  & only \elpr & 102  \\
0.9 & $+0.8$ & $-0.3$  & only \erpl & 120  \\
0.1 & $-0.8$ & $-0.3$  & only \elpl & 359  \\
0.1 & $+0.8$ & $+0.3$  & only \erpr & 396  \\
%\hline
2.0 & \multicolumn{2}{c}{mixed} & full analysis & 75 \\
\end{tabular}
\end{ruledtabular}
\end{table}%%%End of the table

Table~\ref{tab:allresults} compares the expected precisions 
on the \psicp\ measurement, estimated as the median of the distribution of the pseudo-experiments' uncertainties, 
for an integrated luminosity of 1~\invab\ in different scenarios
according to the decay of the Z boson, of the $\tau$ leptons, and the beam polarization.
Events with hadronic Z decays dominate the sensitivity, due to their statistical advantage.
Events in which at least one $\tau$ lepton decays in the $\rho$ channel also dominate, for the same reason.
The beam polarization has only a rather small effect on the precision, with the \elpr\ scenario
slightly favored due to the larger signal cross-section and despite the higher expected backgrounds.

Some sources of non-SM CP effects are expected to also affect the total \zh\ cross-section due to suppression of the ZZH coupling.
The effect of a $\pm 20\%$ variation of the Higgs-strahlung cross-section
(close to the current precision on the measurement of the $\mathrm{H \to ZZ}$ signal strength at LHC)
on the estimated precision is shown in the table. Varying the cross-sections of only processes
with H$\to \tpm$ decays by $\pm 20\%$ results in almost identical sensitivity changes.
%, as are variations in sensitivity induced by an enhancement or suppression of the cross-section of
%processes with H$\to \tpm$ decays.

The table also shows results in which the background is assumed to be completely rejected (``no bg.''),
the $\tau$ lepton polarimeters perfectly measured (``perf. pol.''), and when signal events are selected
with perfect efficiency (``perf. eff.''). The effect of experimental resolution on the polarimeter measurement is rather small, however
the inclusion of realistic backgrounds and signal efficiencies leads to a significant decrease in sensitivity compared to ideal results.
Improved data reconstruction techniques resulting in better separaion between signal and backgrounds
have the potential to significantly enhance the experimental sensitivity.

The expected uncertainty using the 2~\invab\ of 250~GeV data expected at ILC under realistic conditions is
75~mrad. % dec 2017, full bg etc, H20-staged
Data taken with the \elpr\ beam polarization is somewhat more sensitive than \erpl.

\begin{figure}
\centering
\includegraphics[width=0.49\textwidth]{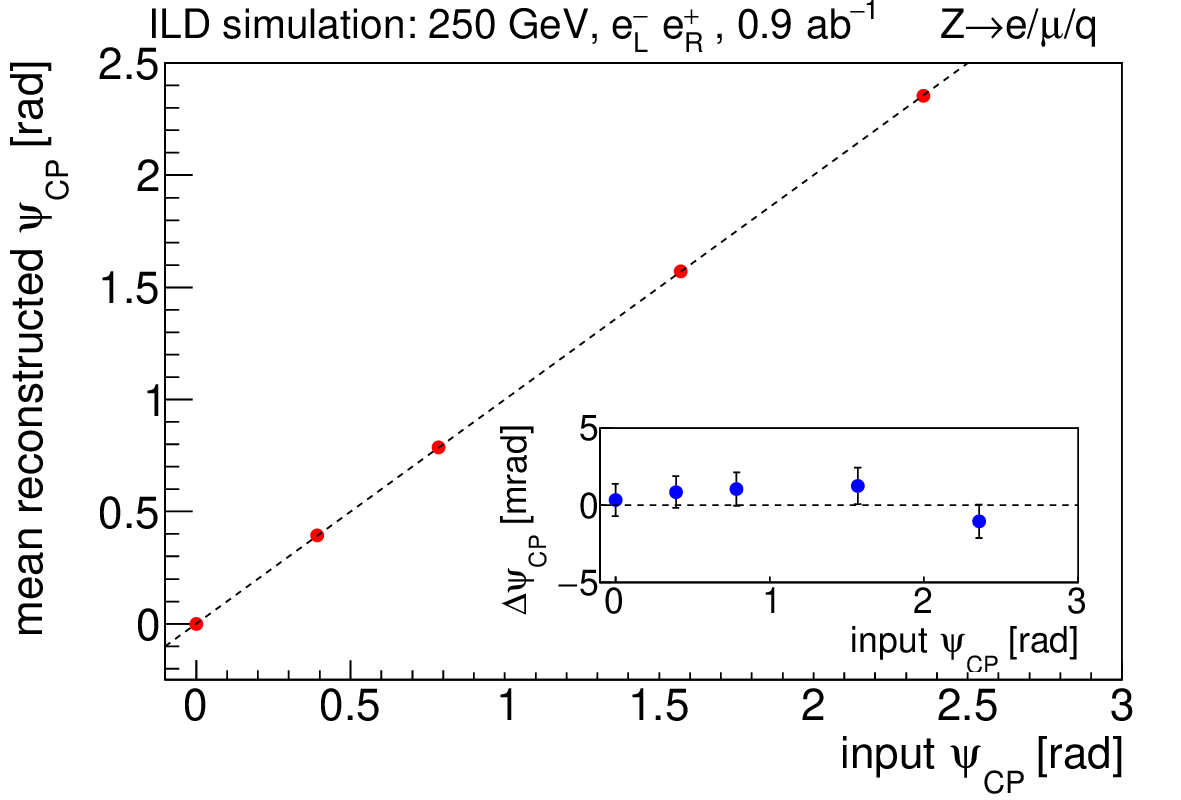}
\caption{
Mean reconstructed \psicp\ values for sets of 10k pseudo-experiments, compared to the true \psicp\ values used in the event generation.
The inset shows the deviations from ideal behavior ($\Delta\psicp$). The dotted lines show ideal behavior.
}
\label{fig:non_sm_psi}
\end{figure}

Figure~\ref{fig:non_sm_psi} shows the dependence of mean extracted \psicp\ values on the value used in the generation of signal simulation samples.
Signal samples with $\psicp = 0$ (i.e. the SM), $\pi/8$, $\pi/4$, $\pi/2$, and $3\pi/4$~rad were used, together with the
usual background samples. The \zh\ cross-section was assumed to be unchanged. 
10k pseudo-experiments were performed for each input \psicp\ phase, assuming the 0.9~\invab\ ILC data at 250~GeV with \elpr\ 
beam polarization. The distributions of the deviations of the extracted \psicp\ phase from the true value 
were each fitted using a Gaussian function, the means of which (and their statistical uncertainties) are shown in the figure.
No systematic bias on the extracted value of \psicp\ due to the fitting method is seen. 
The maximum deviations are of order 1~mrad, consistent with the statistical uncertainties
due to the number of pseudo-experiments, and insignificant compared to the single-experiment sensitivity.

\subsection{Systematic uncertainties}

The main source of systematic uncertainties in this measurement will be due to \Dphi\ dependent signal selection efficiency or
background acceptance.
No such effects were seen in the current analysis, within the fluctuations due to limited simulation statistics.
Significantly larger samples, particularly of background processes, would be required to better constrain the size of any potential effects. 
A well-understood \Dphi\ dependence in either signal or background acceptance can in principle be 
taken into account when extracting the CP phase of the signal.

Large samples of $\mathrm{Z \to \tpm}$ decays from both $\mathrm{e^+ e^- \to Z (\gamma)}$ and $\mathrm{e^+ e^- \to Z Z}$ processes will be produced at ILC, whose
$\tau$ leptons will have similar energies to those from Higgs boson decay used in the present analysis.
Transverse spin correlations between $\tau$ leptons from Higgs decay, which are the bedrock of the present analysis,
are absent in $\tau$ lepton pairs produced in Z or $\gamma$ decays.
These samples will play an essential role in the control of systematic uncertainties in the current analysis, allowing
the simulation, selection and reconstruction of $\tau$ leptons, their decay modes, momenta and polarimeters, 
to be validated in independent event samples.

\section{Conclusions}
\label{sec:conclusion}

The CP nature of the Higgs boson and its couplings are of fundamental importance,
and can be probed in several ways at the ILC and other lepton colliders.
One method, demonstrated in this paper, 
uses the spin correlations of $\tau$ lepton pairs produced in Higgs boson decays.

The Higgs-strahlung process provides a good system
in which to make this measurement. As well as providing a sizable number of events, visible decays of the 
Z boson produced in conjunction with the Higgs boson allow the use of techniques
to fully reconstruct the $\tau$ lepton momenta and polarimeters, and to reject backgrounds.

The correlation between transverse components of the reconstructed $\tau$ polarimeter vectors reflects
the CP nature of the $\tau$-pair. This correlation is stronger in events 
in which the longitudinal polarimeter configuration gives a large contrast function.

Events were selected in three channels, according to the Z boson decay: electrons, muons, and hadrons.
Distinct selection strategies were developed for the leptonic and hadronic channels due to their
different characteristics. Total reconstruction and selection efficiencies range from 
33\% for $\mathrm{Z\to bb}$ events to 
47\%  %v34
in $\mathrm{Z\to \mu\mu}$, and the signal purity of the selected samples varies between 10 and 30\%.
The selection efficiency does not depend significantly on the CP-sensitive observables used
in the analysis, and does not introduce a bias to the measurement.
Event-by-event sensitivity to CP effects was estimated by means of the significance of the impact parameter 
measurement, the reconstructed contrast function, and the output of two neural networks trained to separate signal from background events.

Pseudo-experiments were used to estimate the sensitivity to the CP mixing angle \psicp\ achievable 
using 2~\invab\ of ILC data at 250 GeV, resulting in an expected error of 
75~mrad  %v34
when combining all three selection channels. 
The greatest sensitivity is provided by the hadronic Z decay channel, which benefits from larger statistics than the 
leptonic channels, although the reconstruction quality and background levels are somewhat worse.
This underlines the importance of excellent hadronic jet reconstruction and energy measurement at ILC, one of the 
central motivations for the design of the ILD detector.

The present analysis uses only $\taupi$ and $\taurho$ decays, corresponding to around 37\% of $\tau$ decays,
or 14\% of \Htpm\ events.
In principle, all hadronic $\tau$ lepton decays have equal analysing power in their polarimeters~\cite{KuhnHadronic},
however for other modes the polarimeters can be less trivial to extract, 
and the final states may be more difficult to completely reconstruct.
If these additional $\tau$ lepton decays can be utilized, the fraction of useful \Htpm\ events would increase from 14\% to 42\%,
potentially resulting in a significant increase in sensitivity.
Leptonic $\tau$ lepton decays present a less well constrained system, due to the production of two neutrinos per decay, 
requiring the use of additional constraints to fully reconstruct $\tau$ lepton momenta and polarimeters; the power
of their reconstructable polarimeters is also intrinsically less than for hadronic decays.

It is foreseen that the ILC will also deliver an integrated luminosity of 4~\invab\ at 500~GeV after an energy upgrade~\cite{ILC250phys}.
Although the Higgs-strahlung cross-section at 500~GeV is suppressed by a factor of around three compared to 250~GeV,
a similar analysis can be applied to these events. Some aspects of event reconstruction will be simpler at the higher energy
(for example, better jet energy resolution and a larger average separation between the hadronic and $\tau$ systems due to larger
boosts of the H and Z), and others more difficult (for example worse lepton momentum resolution, and narrower $\tau$ jets).

Higgs boson production in conjunction with a neutrino pair can occur via both the Higgs-strahlung and WW-fusion processes.
The former is significant at 250~GeV, due to the large branching ratio of the Z to neutrinos, while
the latter process becomes important at center-of-mass energies above around 350~GeV.
Such events are however more difficult to reconstruct, due the presence of two additional neutrinos in the 
final state and limited information about the interaction point on an event-by-event basis. 
It it likely that appropriate methods can be developed to deal with these aspects of 
event reconstruction, particularly in the case of multi-prong $\tau$ lepton decays, and that 
additional sensitivity to the Higgs boson CP properties in $\tau$ lepton decays can be obtained from such final states.

\begin{acknowledgments}                                                                                                                                                                                   

We thank the LCC generator working group and the ILD software working group for providing the 
simulation and reconstruction tools and producing the Monte Carlo samples used in this study,
and 
% G.~Wilson, 
K.~Desch, S.~Komamiya, and J.~Yan for their useful comments on drafts of this paper.
This work has benefited from computing services provided by the ILC Virtual Organization, 
supported by the national resource providers of the EGI Federation and the Open Science GRID.

\end{acknowledgments}

% \bibliography{my}


\end{document}